\documentclass[prd, twocolumn, amsmath, nofootinbib]{revtex4-2}

\usepackage{graphicx}
\usepackage{dcolumn}
\usepackage{bm}
\usepackage{booktabs}
\usepackage{multirow}
\usepackage{siunitx}
\usepackage{orcidlink}
\usepackage{placeins}
\usepackage{subfig }
\usepackage{url}
\newcommand{\pulsarpanel}[2]{%
  \begin{minipage}[t]{0.32\textwidth}
    \centering
    \includegraphics[width=\linewidth]{#1}\par
    \vspace{-2pt}
    {\scriptsize #2}
    \vspace{4pt}
  \end{minipage}%
}

\begin{document}

\preprint{APS/123-QED}

\title{Insights into the Pulsar Timing Array hypothesis space}

\newcommand{\AUTMaths}{Department of Mathematical Sciences, Auckland University of Technology, Private Bag 92006, Auckland 1142, New Zealand}
\newcommand{\Manly}{Manly Astrophysics, 15/41-42 East Esplanade, Manly, NSW 2095, Australia}
\newcommand{\UoAStats}{Department of Statistics, University of Auckland, 38 Princes St, Auckland, New Zealand}

\author{El Mehdi Zahraoui$^{1}$\orcidlink{0009-0006-0900-3824}}
\author{Patricio Maturana-Russel$^{1,2}$\orcidlink{0000-0002-5211-9818}}
\author{Willem van Straten$^{3}$\orcidlink{0000-0003-2519-7375}}
\author{Renate Meyer$^{2}$\orcidlink{0000-0003-0268-8569}}
\author{Sergei Gulyaev$^{1}$\orcidlink{0000-0003-0186-5551}}

\affiliation{$^{1}$\AUTMaths}
\affiliation{$^{2}$\UoAStats}%
\affiliation{$^{3}$\Manly}

\date{\today}

\begin{abstract}
We present novel insights into the pulsar-noise model space in pulsar timing array (PTA) experiments. Through a comparative analysis of the same Parkes PTA second data release observations processed with the 2020 and 2023 pipelines, we show that data processing materially changes the distribution of  posterior support across competing noise hypotheses and increases sensitivity to weak contributions such as the solar wind. A solar-wind component appears in the highest-posterior hypothesis for ten pulsars under the 2023 pipeline, against four under the 2020 pipeline, while the corresponding mean electron density estimates are consistent with PPTA DR3 results. Furthermore, hypothesis-space analysis exposes model competition and degeneracies that are hidden by single-model summaries. Approximately 75\% of the pulsars retain substantial posterior support for alternative noise descriptions. Therefore, understanding the nature of the pulsar noise hypothesis space is crucial  for robust inference and nanohertz gravitational-wave background searches.

\end{abstract}

\maketitle


\section{Introduction}

Pulsar timing arrays (PTAs) were conceived as Galactic-scale detectors of gravitational waves (GWs) in the nanohertz frequency band \cite{sazhin1978opportunities,hellings1983upper}. By monitoring the pulse times of arrival from an ensemble of exceptionally stable millisecond pulsars (MSPs), PTAs search for the characteristic quadrupolar angular correlations induced by an isotropic stochastic gravitational-wave background (GWB) in the pulsar timing residuals \cite{Foster1990}. Over the past several decades, major collaborations including the North American Nanohertz Observatory for Gravitational Waves \cite[NANOGrav;][]{McLaughlin_2013}, the Parkes Pulsar Timing Array \cite[PPTA;][]{manchester2013parkes}, the European Pulsar Timing Array \cite[EPTA;][]{ferdman2010european}, the Indian Pulsar Timing Array \cite[InPTA;][]{joshi2018precision}, the MeerKAT Pulsar Timing Array \cite[MeerKAT PTA;][]{Meerkat2023}, and the Chinese Pulsar Timing Array \cite[CPTA;][]{CPTA}, have assembled increasingly longer and more precise pulsar-timing data sets. 

Through Bayesian analyses, recent data releases have provided strong evidence for a common red noise spectrum process, with several analyses also reporting inter-pulsar spatial correlations consistent with those expected from a GWB \cite{reardon2023gravitational,agazie2023nanograv,antoniadis2023second,zic2023parkes,miles2023meerkat}. However, the timing residuals entering these analyses are themselves products of calibration, radio-frequency-interference excision, frequency averaging, pulse-profile template construction, and time-of-arrival estimation \citep{zic2023parkes}. These processing choices are not statistically neutral: improved instrumental calibration and template matching can reduce apparent white noise and alter the inferred properties of red noise \cite{rogers2024reducing}, while inadequately modelled frequency-dependent profile evolution and shape variability can bias measured arrival times \cite{pennucci2019frequency,lentati2017wide}. Because the adopted single-pulsar noise models can affect the inferred properties of a common red-spectrum process \cite{reardon2023gravitational}, processing-induced changes in the timing residuals may propagate into PTA-wide gravitational-wave conclusions.


Bayesian inference has become a central component of PTA data analysis \cite{van2009measuring,Lentati2013PhRvD,Ellis2013}. It is widely used both to characterize stochastic processes affecting individual MSPs \cite{goncharov2021identifying,agazie2023nanograv,antoniadis2023second,zic2023parkes,chen2025chinese} and to perform joint searches for a common GWB signal \cite{reardon2023gravitational,epta2024second,agazie2023nanograv}. A robust assessment of the evidence for a GWB requires careful treatment of the stochastic processes that can affect PTA sensitivity \cite{2016MNRAS.458.2161L}. These include intrinsic pulsar spin noise, commonly referred to as red noise (RN), variations in the interstellar dispersion measure (DM), and dispersive contributions from the solar wind (SW) \cite{2016MNRAS.458.2161L,goncharov2021identifying,hazboun_bayesian_2022}. Different combinations and parameterizations of these components give rise to a large discrete space of candidate noise models. Consequently, several studies have used Bayesian model comparison to identify preferred noise descriptions for individual pulsars and PTA data sets \cite{jones2017nanograv,goncharov2021identifying,2022MNRAS.509.5538C,antoniadis2023second2}.

To navigate this large model space, PTA analyses have increasingly adopted model-index and trans-dimensional sampling methods. Product-space sampling, often referred to in PTA analyses as the hypermodel approach, embeds a collection of candidate models within a common augmented parameter space and introduces a discrete model-index parameter that determines which model is active \cite{carlin1995bayesian,ellis2019enterprise,justin_ellis_2017_1037579}. More recently, inference frameworks such as \texttt{tPTAbilby} have introduced binary indicator variables for candidate stochastic processes, allowing the number and combination of active noise components to be inferred directly from the data \cite{dimarco2026transdimensionalsamplingframeworkpulsar}. These approaches permit parameter estimation and Bayesian model selection to be performed simultaneously within a single sampling run, while naturally incorporating the Bayesian penalty for unnecessary model complexity. Provided that the sampler mixes efficiently between models, posterior model probabilities can be estimated from their relative occupancy fractions, and Bayes factors can be obtained from ratios of these probabilities if the models are assumed apriori equally probable  \cite{justin_ellis_2017_1037579,dimarco2026transdimensionalsamplingframeworkpulsar}.

Recent advances in PTA marginal-likelihood estimation, including generalized steppingstone sampling (GSS) \citep{zahraoui2025generalized} and MorphZ \citep{morphz}, have substantially improved the feasibility of estimating marginal likelihoods and, in turn, posterior model probabilities. Unlike product-space and trans-dimensional methods, which estimate model probabilities from the posterior occupancy of competing configurations \citep{carlin1995bayesian,dimarco2026transdimensionalsamplingframeworkpulsar}, these approaches evaluate the evidence for each model independently allowing a direct estimation of  posterior model probabilities and Bayes factors \citep{kass1995bayes}. This enables parallel computation and model-specific convergence checks, and allows evidences to be reused when the candidate set or model priors change. Moreover, their efficiency extends to high-dimensional PTA models, including searches for common processes \citep{zahraoui2025generalized,morphz}.

Here, we use MorphZ to compare a fixed set of pulsar-noise models across the original PPTA DR2 \cite{pptadr2_2021} and its 2023 DR3 reprocessing \cite{pptadr3_2023}. By examining the posterior model probabilities, we identify competing hypotheses, parameter degeneracies, and cases where the data cannot clearly distinguish between different combination of noise components. We then assess how changes in calibration, radio-frequency-interference mitigation, time-of-arrival estimation, and observing-system treatment affect inferred model preferences. This comparison reveals which noise-model conclusions are robust to reprocessing and which depend sensitively on the construction of the timing data set.

This paper is organized as follows. Section~\ref{sec:bayesian_inference} introduces the Bayesian model comparison framework and defines the posterior model probabilities used throughout the analysis. Section~\ref{sec:data} describes the original and reprocessed PPTA DR2 data sets. Section~\ref{sec:analysis} presents the pulsar-noise models, prior choices, and marginal-likelihood calculations. Section~\ref{sec:results} compares the resulting model spaces and examines the impact of data reprocessing on noise-model preferences. Finally, Section~\ref{sec:conclusion} summarizes the main findings and their implications for PTA gravitational-wave searches.

\section{Bayesian inference}
\label{sec:bayesian_inference}

For a dataset \(D\) and a finite set of $K$ competing hypotheses \(\mathcal{H}=\{H_1,\ldots,H_K\}\), the posterior probability of
hypothesis \(H_i\) is given by
\begin{align}
    p(H_i\mid D,\mathcal{H})
    &=
    \frac{p(D\mid H_i)\,p(H_i\mid\mathcal{H})}
    {\displaystyle\sum_{j=1}^{K}
    p(D\mid H_j)\,p(H_j\mid\mathcal{H})},
    \label{eq:model_space}
\end{align}
where \(p(H_i\mid\mathcal{H})\) is the prior probability assigned to
\(H_i\), and
\begin{align}
    Z_i \equiv p(D\mid H_i)
    =
    \int_\Theta
    p(D\mid\boldsymbol{\theta}_i,H_i)\,
    p(\boldsymbol{\theta}_i\mid H_i)\,
    \mathrm{d}\boldsymbol{\theta}_i,
    \label{eq:marginal_likelihood}
\end{align}
where $\boldsymbol{\theta}_i \in \Theta_i$ is the parameter vector for model $H_i$, is its marginal likelihood, or evidence. The denominator of
Eq.~\eqref{eq:model_space},
\begin{align}
    p(D\mid\mathcal{H})
    =
    \sum_{j=1}^{K} Z_j\,p(H_j\mid\mathcal{H}),
    \label{eq:model_space_evidence}
\end{align}
is the model space normalizing constant. Because each fixed-model evidence \(Z_j\) is estimated separately, this normalizing constant is obtained without additional computation, enabling the estimation of posterior model probabilities.

The model priors  \(p(H_i\mid\mathcal{H})\) may represent different expectations about the occurrence of
intrinsic red noise, dispersion-measure variations, solar-wind variations, or additional white-noise components. Uniform model priors assign equal prior support to every candidate model, whereas informed
priors can favor simpler models or increase the prior probability of physically motivated noise components. Posterior model probabilities may therefore be recomputed under different model-prior choices without extra sampling cost, provided
that \(p(D\mid\boldsymbol{\theta}_i,H_i)\) and \(p(\boldsymbol{\theta}_i\mid H_i)\) remain unchanged. 

\begin{table}
\centering
\caption{Priors definition of the noise components used in the noise models for the PTA DR2 analysis.}
\label{tab:noise_components}
\renewcommand{\arraystretch}{1.3} 
\begin{tabular}{|c|c|c|}
\hline
\textbf{Noise component}& \textbf{Parameters} & \textbf{Priors} \\ \hline
\multirow{3}{*}{WN}& EFAC & $\mathcal{U}(0.01, 10)$ \\ \cline{2-3} 
 & EQUAD & $\log_{10} \mathcal{U}(10^{-9}, 10^{-5})$ \\ \cline{2-3} 
 & ECORR & $\log_{10} \mathcal{U}(10^{-9}, 10^{-2})$ \\ \hline
\multirow{2}{*}{RN}& $A_{\text{RN}}$ & $\log_{10} \mathcal{U}(10^{-20}, 10^{-11})$ \\ \cline{2-3} 
 & $\gamma_{\text{RN}}$ & $\mathcal{U}(0, 7)$ \\ \hline
\multirow{2}{*}{DMv}& $A_{\text{DMv}}$ & $\log_{10} \mathcal{U}(10^{-20}, 10^{-11})$ \\ \cline{2-3} 
 & $\gamma_{\text{DMv}}$ & $\mathcal{U}(0, 7)$ \\ \hline
\multirow{3}{*}{SW}& $n_{\text{earth}}$ & $\mathcal{U}(0, 30)$ \\ \cline{2-3} 
 & $A_{\text{SW}}$ & $\log_{10} \mathcal{U}(10^{-11}, 10^1)$ \\ \cline{2-3} 
 & $\gamma_{\text{SW}}$ & $\mathcal{U}(-2, -1)$ \\ \hline
\end{tabular}
\end{table}

\section{Data}
\label{sec:data}
In this study, we analyze two instances of the Parkes Pulsar Timing Array (PPTA) second data release to assess the sensitivity of single-pulsar noise models to data processing choices prerequisite to the Bayesian analysis phase. The first is the original PPTA-DR2 \cite{kerr2020parkes}, as used in the timing analysis of \cite{pptadr2_2021}. The second is the reprocessed version of the same observations, produced as part of the PPTA third data release \cite{zic2023parkes}. Both datasets share a common set of 25 millisecond pulsars observed with the 64-m Parkes radio telescope. 
\begin{figure*}
    \centering
    \includegraphics[width=\linewidth]{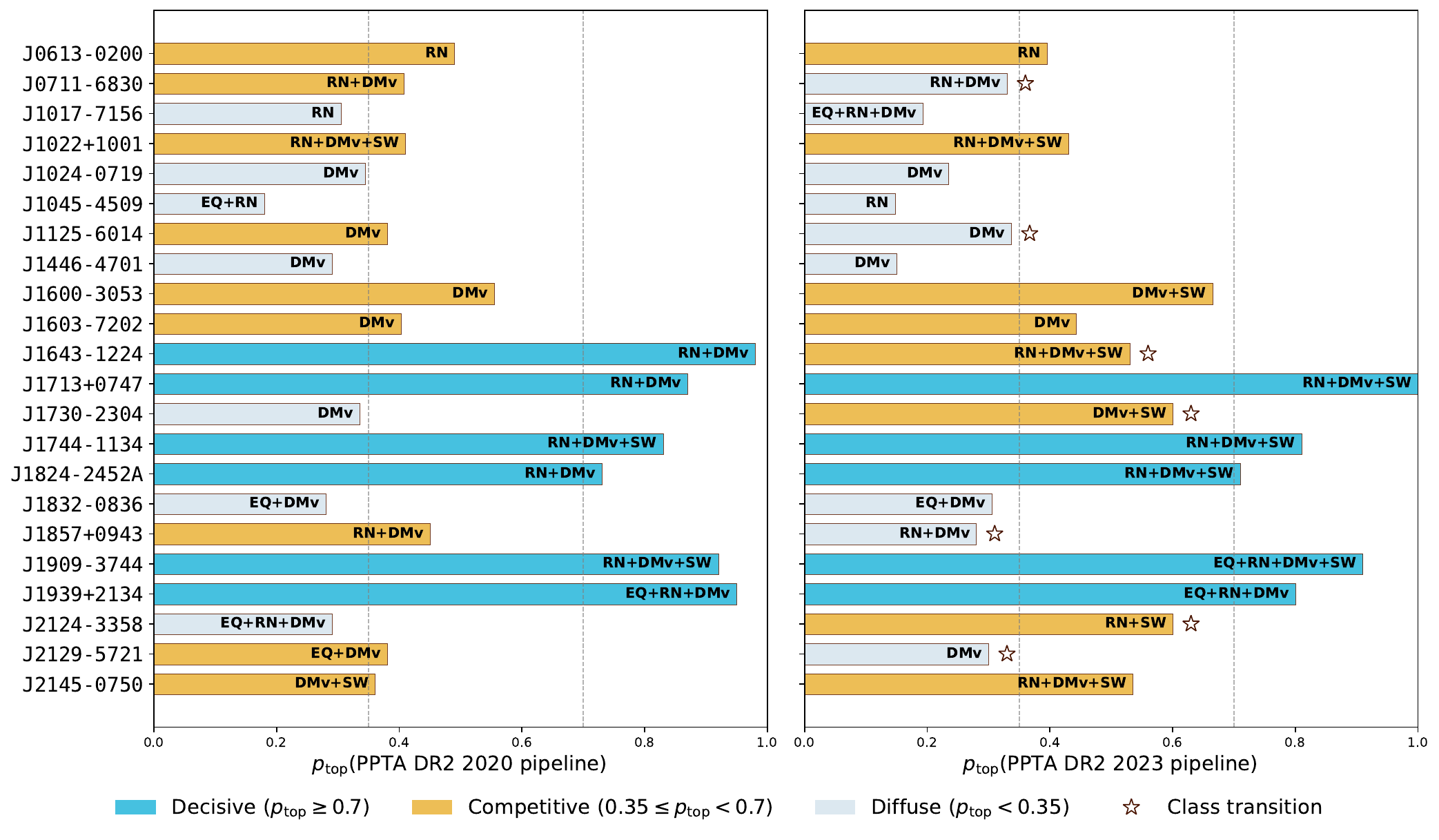}
    \caption{Summary of the model space for 22 pulsars noise analysis. For each pulsar, the model with the highest posterior probability $p_{\text{top}}$ is shown for both the 2020~DR2 (left) and 2023~DR2 (right) pipelines.
    Bars are colour-coded by class: decisive ($p_{\text{top}}\ge0.7$; cyan), competitive ($0.35\le p_{\text{top}}<0.7$; light orange), and diffuse ($p_{\text{top}}<0.35$; light grey).
    Abbreviated top models are annotated; EF and EC are present in every $p_{\text{top}}$ model and omitted for clarity. Stars mark pulsars that change classes between datasets.}
    \label{fig:class_bars}
\end{figure*}

The original DR2 provides sub-banded arrival times derived from calibrated pulse profiles with 32 frequency channels per observation, dynamically sub-banded to maintain a minimum signal-to-noise ratio \cite{kerr2020parkes}. Times of arrival (ToAs) were obtained by cross-correlating profiles against band-specific analytic templates. The timing models and noise characterization are described in \cite{pptadr2_2021} and \cite{goncharov2021identifying}, respectively. The data are referred to the TT(BIPM2018) timescale and the JPL DE436 solar system ephemeris.  

The reprocessed DR2 was produced to enable consistent combination with new ultra-wide-bandwidth (UWL)  receiver observations \cite{pptadr3_2023}. The reprocessing involved several changes, including improved radio-frequency interference excision using MeerGuard \cite{lazarus2016prospects}, reduction to four frequency channels per band, new ToA templates derived from UWL wide-band pulse portraits \cite{curylo2023pulse}, Fourier-domain Monte Carlo ToA estimation via pat \cite{hotan2004psrchive}, updated dispersion measure values and timing ephemerides, and adoption of the TT(BIPM2020) reference timescale. Additional frequency-dependent (FD) parameters were also required to absorb residual profile-evolution systematics introduced by the wide-band portrait templates. The full details of these changes are described in the DR3 release paper \cite{zic2023parkes}. In this work, we restrict our analysis to the 22 pulsars common to both datasets, enabling a direct comparison of the inferred noise models.

\section{Analysis}
\label{sec:analysis} 
In the following, we refer to the different hypotheses as noise models. We construct the noise models using \textsc{Enterprise}~\cite{ellis2019enterprise,enterprise} via \textsc{Enterprise Warp}~\footnote{\textsc{Enterprise Warp}: \url{https://enterprise-warp.readthedocs.io/}}.
Posteriors are sampled with \textsc{emcee}~\cite{emcee} and cross-checked with \textsc{PocoMC}~\cite{karamanis2022accelerating,karamanis2022pocomc}; all marginal likelihoods are estimated with \textsc{MorphZ}~\cite{morphz}. The posterior model probabilities are computed directly using these marginal likelihood estimates and a custom prior model probabilities.

We adopt the following notation: EF for EFAC (error scale factor),
EQ for EQUAD (error added in quadrature), and EC for ECORR (extra
correlated noise) as the white-noise parameters; and RN,
$\mathrm{DM_{v}}$, and SW for achromatic red noise,
dispersion-measure variations, and solar-wind variations,
respectively\cite{goncharov2021identifying}. Each stochastic process (SP) is modeled as a power law with two free parameters: the log-amplitude $\log_{10}A_\text{SP}$ and a spectral index $\gamma_{\text{SP}}$. We use a SW model which includes an additional parameter: $n_{\mathrm{earth}}$ the mean free electron density at 1 AU \cite{reardon2023gravitational,hazboun_bayesian_2022,iraci2025combining}. Table~\ref{tab:noise_components} shows the prior distribution used for each noise component parameters.

For each pulsar and dataset, we evaluate K=13 candidate noise models after excluding white-noise only models. These excluded models have a zero posterior model probabilities for every pulsar in both datasets, due to a presence of excess noise in the dataset. These models span different combinations of EQ, EC, RN, and DMv. For the solar wind, we present only the four following models: EC+RN+SW, EC+DMv+SW, EC+RN+DMv+SW, and EQ+EC+RN+DMv+SW. These models are added to determine if the solar wind contribution is present in the dataset. For each pulsar, the thirteen models  are shown in Figure~\ref{fig:model_probs} and in the appendix~\ref{app:model_probabilities} ( Figure~\ref{fig:app_model_1} and \ref{fig:app_model_2}), where EF is included in all models and is omitted from abbreviated labels. Deterministic signals and chromatic noise are included as free components for each relevant pulsar, following the description in  \citeauthor{goncharov2021identifying}\cite{goncharov2021identifying}. For each stochastic component, the number of Fourier harmonics is set to 30.  

Our approach differs from the sequential procedure of Ref.~\cite{goncharov2021identifying}, where white-noise parameters are first fixed at their maximum-posterior values before red-noise components are selected iteratively. Such fixing influences the inference when white-noise and red-noise parameters are covariant \citep{goncharov2021identifying}. In contrast, we evaluate 13 noise-component combinations simultaneously while keeping the white-noise parameters free, allowing the data rather than the procedure to determine the preferred model.

\section{Results}
\label{sec:results}

We compare the candidate models through their posterior model probabilities defined in Eq.~\eqref{eq:model_space}, assuming uniform prior probabilities over the $K=13$ models considered for
each pulsar. Throughout this section, the posterior probability of the highest-probability model is denoted by $p_{\rm top}$. This quantity provides a compact description of the concentration of the model posterior, but it does not contain all of the information in the model space. In particular, two pulsars with the same value of
$p_{\rm top}$ may differ substantially in how the remaining posterior probability is distributed among the other models.
\begin{figure}
    \centering
    \includegraphics[width=1\columnwidth]
        {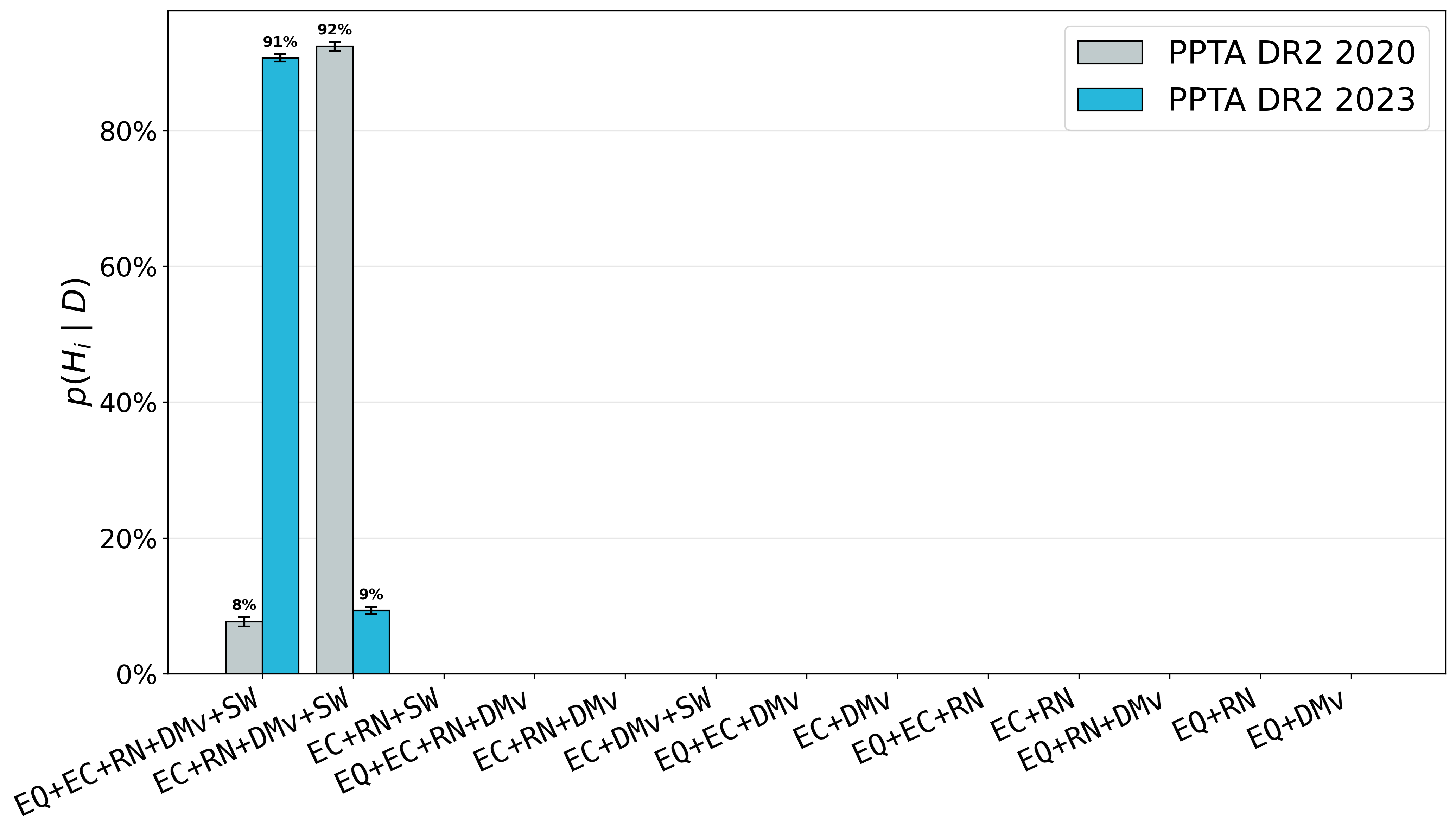}

    \vspace{0.5em}
    {\small (a) PSR~J1909$-$3744}

    \vspace{1em}

    \includegraphics[width=1\columnwidth]
        {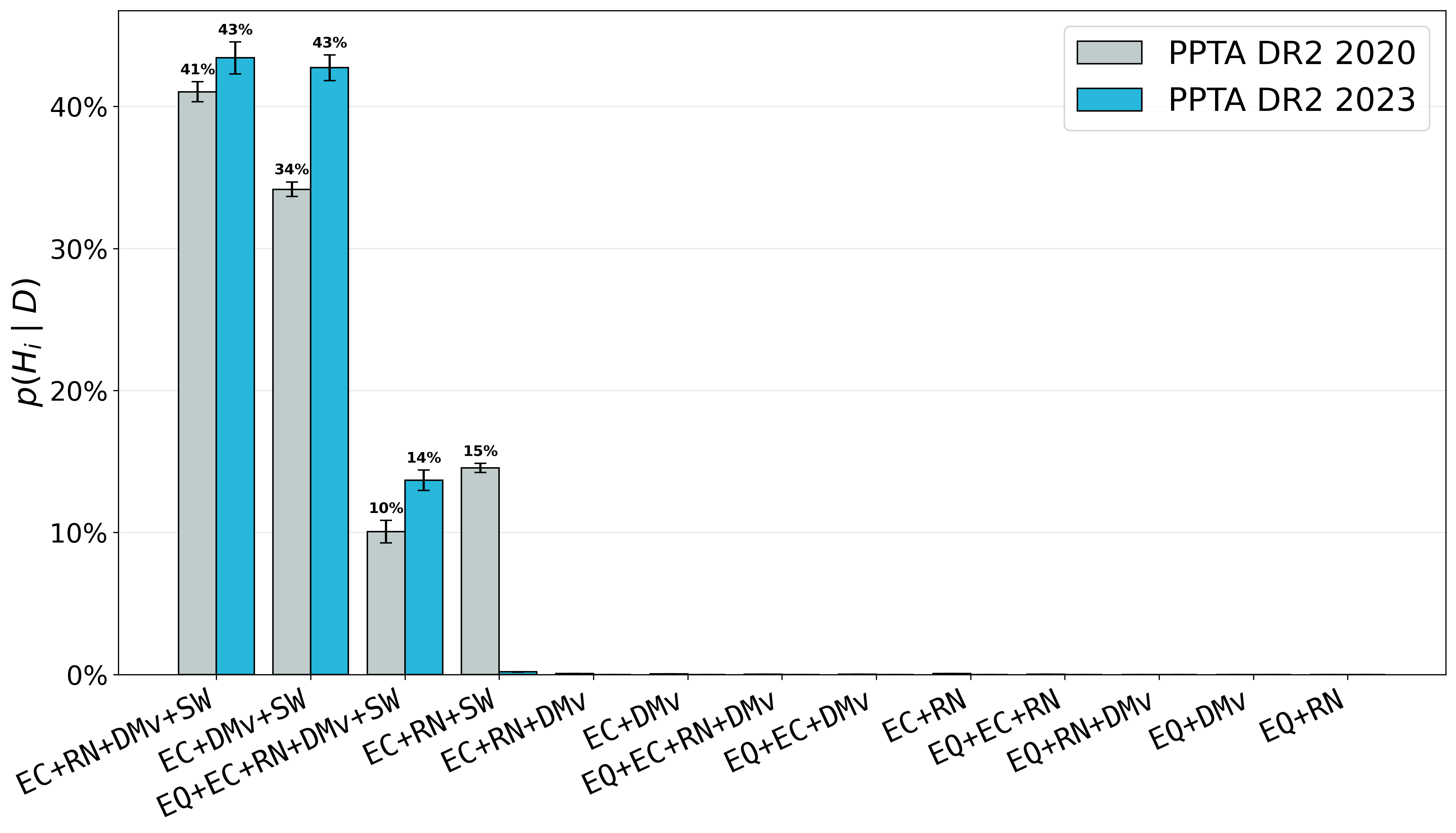}

    \vspace{0.5em}
    {\small (b) PSR~J1022$+$1001}

    \vspace{1em}

    \includegraphics[width=1\columnwidth]
        {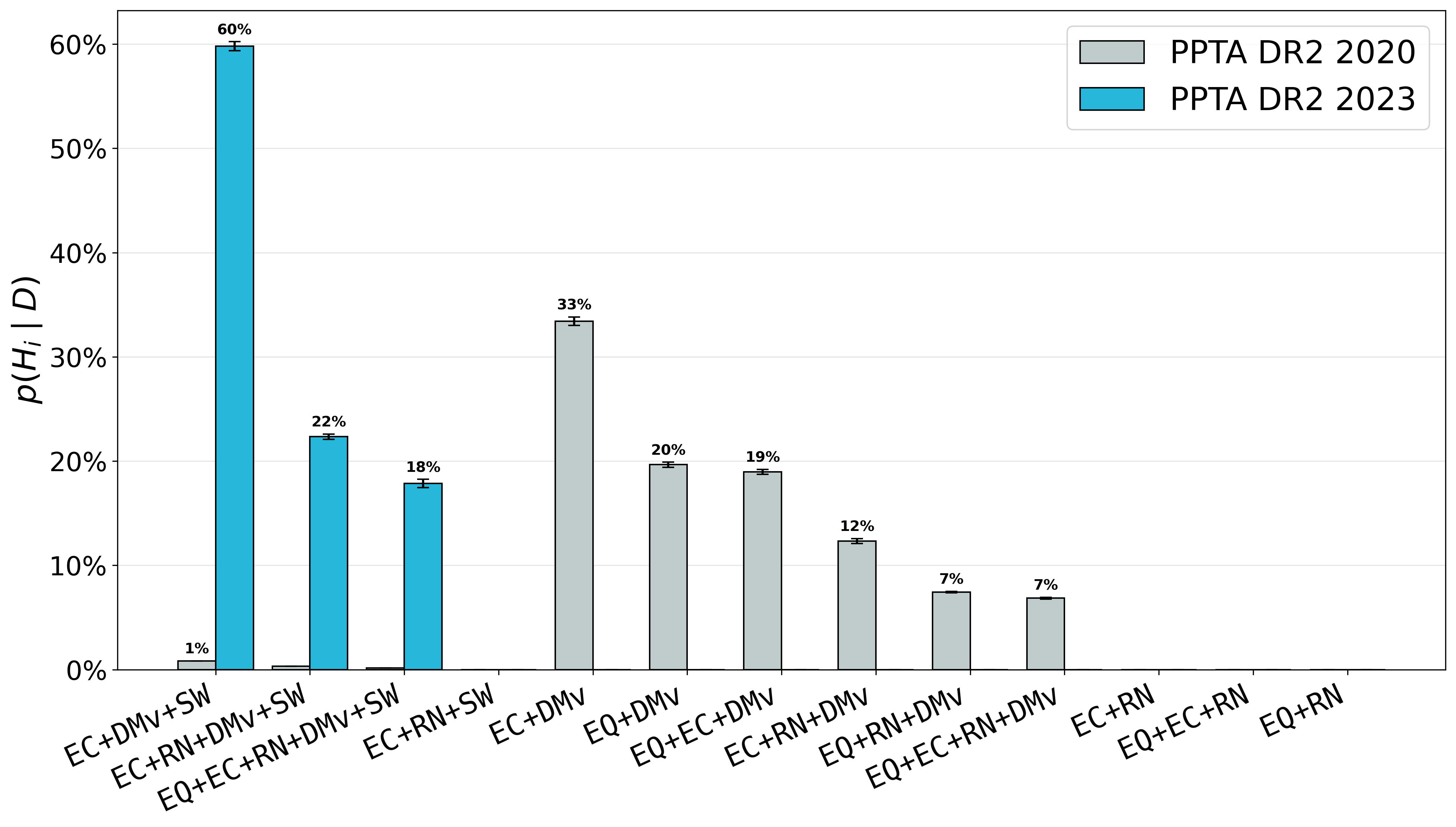}

    \vspace{0.5em}
    {\small (c) PSR~J1730$-$2304}
    \caption{Posterior model probabilities for one representative pulsar from each class under the 2020~DR2 (grey) and 2023~DR2 (cyan) pipelines.
    \textit{(a)}~Decisive: one model dominates at ${\sim}\,90\%$.
    \textit{(b)}~Competitive: two SW-containing models each hold ${\sim}\,40\%$.
    \textit{(c)}~Diffuse: the posterior fragments in 2020 but reorganizes around DMv+SW in 2023.}
    \label{fig:model_probs}
\end{figure}

In the following, a pulsar is labeled \emph{decisive}
when $p_{\rm top}\geq0.7$, \emph{competitive} when
$0.35\leq p_{\rm top}<0.7$, and \emph{diffuse} when
$p_{\rm top}<0.35$. We use this classification, as shown in Figure~\ref{fig:class_bars},
only as a descriptive summary of the model spaces of these datasets. These arbitrary thresholds are not used to accept the highest-probability model or to discard the remaining models. Instead,
they distinguish model spaces dominated by one model from those
in which the data retain substantial support for multiple noise
descriptions.

\subsection{Precision of the marginal-likelihood estimates}

First, we assess the
precision of the marginal-likelihood estimates used to construct
the posterior model probabilities. Table~\ref{tab:logz_preci}
summarizes the run-to-run variation across five independent
calculations for all model--pulsar combinations. For each dataset,
the summary contains 1430 $\log(z)$ estimates. The
median standard deviation across repeated calculations is $0.016$ for the 2020 DR2 analysis and $0.013$ for the 2023 DR2 analysis. The
corresponding coefficient of variation values are {$9.23\cdot10^{-5}$} and {$1.78\cdot10^{-4}$},
respectively, indicating a low variance between the independent runs.
\begin{table}[t]
\centering
\caption{ Summary of the log marginal-likelihood estimates
for the DR2 2020 and DR2 2023 datasets. The median and standard deviation of $\sigma_{\log(\widehat{z})}$ the standard deviation of $5\times286$
independent evidence estimates. The averaged coefficient of variation \(\overline{\text{CV}}_{\log(\widehat{z})}\) across these evidence estimates. }

\begin{tabular}{
    l
    S[table-format=3.0]
    S[table-format=1.5]
    S[table-format=1.5]
    S[table-format=1.5]
}
\toprule
{Dataset}
& {\(N_{{\log(\widehat{z})}}\)}
& {\(\operatorname{med}_{\sigma_{\log(\widehat{z})}}\)}
& {\(\operatorname{std}_{\sigma_{\log(\widehat{z})}}\)}
& {\(\overline{\text{CV}}_{\log(\widehat{z})}\)}
\\
\midrule
DR2 2020 & 1430 & 0.016 & 0.091 & {$9.23\cdot10^{-5}$}  \\
DR2 2023 & 1430 & 0.013 & 0.058 & {$1.78\cdot10^{-4}$} \\
\bottomrule
\label{tab:logz_preci}
\end{tabular}
\end{table}
The close agreement between independent runs demonstrates that the marginal-likelihood estimates are numerically stable for both datasets. This stability is essential for the next stage of the analysis, because it allows the uncertainty in the marginal-likelihood estimates to be propagated reliably into the posterior model probabilities shown in the figures. 

The model probabilities uncertainties (see Figure~\ref{fig:model_probs}) were obtained by a first-order propagation of the log-evidence uncertainties through the normalized exponential transformation. The log-evidence estimates are assumed to be independent, and model-prior probabilities were treated as fixed. Consequently, the uncertainty in each model probability includes contributions from the log-evidence uncertainties of all models entering the normalization, as described in Eq.~\ref{eq:model_space}.
Consequently, the differences observed between competing models and between processing pipelines are not dominated by numerical fluctuations in the evidence calculations, but instead reflect genuine changes in the support provided by the data.  

\subsection{Summary of PPTA DR2 model space}

Figure~\ref{fig:class_bars} summarizes the model spaces of the
22 pulsars through $p_{\rm top}$ and the composition of the
highest-probability model. Under the 2020 pipeline, six pulsars are
classified as decisive, nine as competitive, and seven as diffuse.
Under the 2023 pipeline, the corresponding numbers are five, eight,
and nine. Thus, neither dataset generally identifies a unique noise
model. The competitive and diffuse classes contain the majority of
the pulsars under both processing pipelines.

Seven pulsars change classification between datasets. These
transitions occur in both directions. PSRs J0711$-$6830,
J1125$-$6014, J1857+0943, and J2129$-$5721 move from the competitive
class in 2020 to the diffuse class in 2023. For these pulsars, the
reprocessed data distribute the posterior probability more broadly
among the candidate models. PSR J1643$-$1224 moves from decisive to
competitive, because its previously dominant model loses posterior
probability to the SW variant models after reprocessing. Conversely,
PSRs J1730$-$2304 and J2124$-$3358 move from diffuse to competitive,
showing a concentration of posterior support for multiple SW models in the 2023 analysis.

The presence of transitions in both directions shows that improved reprocessing does not make every model posterior
more concentrated. Rather, the changes in calibration, interference
mitigation, template construction, ToA estimation, and timing-model
treatment alter the relative ability of the candidate stochastic
processes to describe each pulsar. For some pulsars this resolves
part of the model ambiguity, whereas for others it exposes additional
competition that was less apparent in the original dataset.

The most prominent change with the reprocessed data concerns the solar wind contribution. When models differing only in their white-noise components are grouped into the same stochastic process family, the most common highest probability family changes from RN+DMv in the 2020 analysis to RN+DMv+SW in the 2023 analysis. A solar wind component occurs in the highest-probability model for four pulsars under the original
pipeline and for ten pulsars under the reprocessed pipeline. The ten
pulsars are PSRs J1022+1001, J1600$-$3053, J1643$-$1224,
J1713+0747, J1730$-$2304, J1744$-$1134, J1824$-$2452A,
J1909$-$3744, J2124$-$3358, and J2145$-$0750. For these ten pulsars, Table~\ref{tab:sw_comparison} reports solar-wind-density estimates from the highest-posterior SW-containing model under the 2023 pipeline.

\begin{figure}
    \centering

    \includegraphics[width=0.72\columnwidth]
        {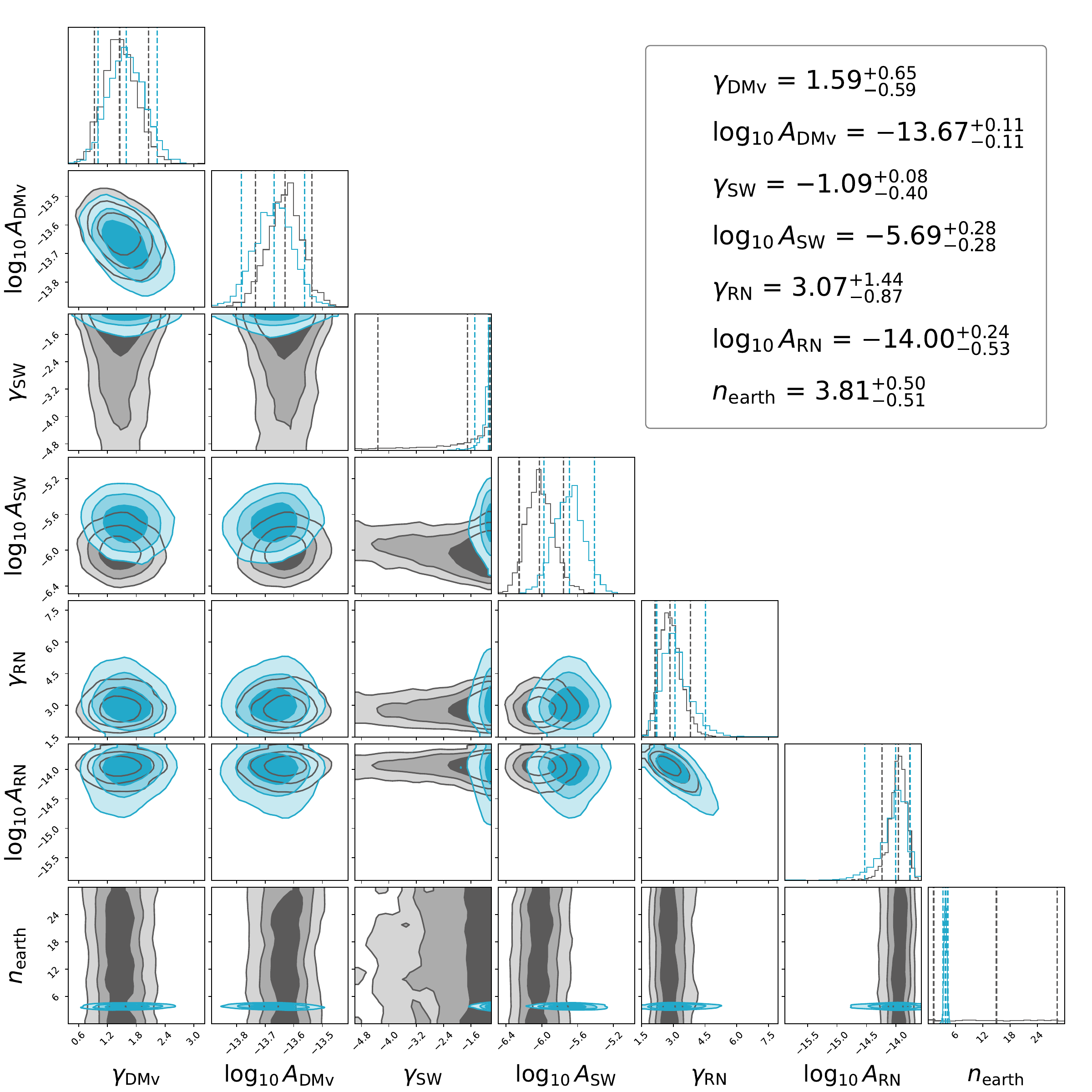}

    \vspace{0.1em}
    {\small (a) PSR~J1909$-$3744}


    \includegraphics[width=0.72\columnwidth]
        {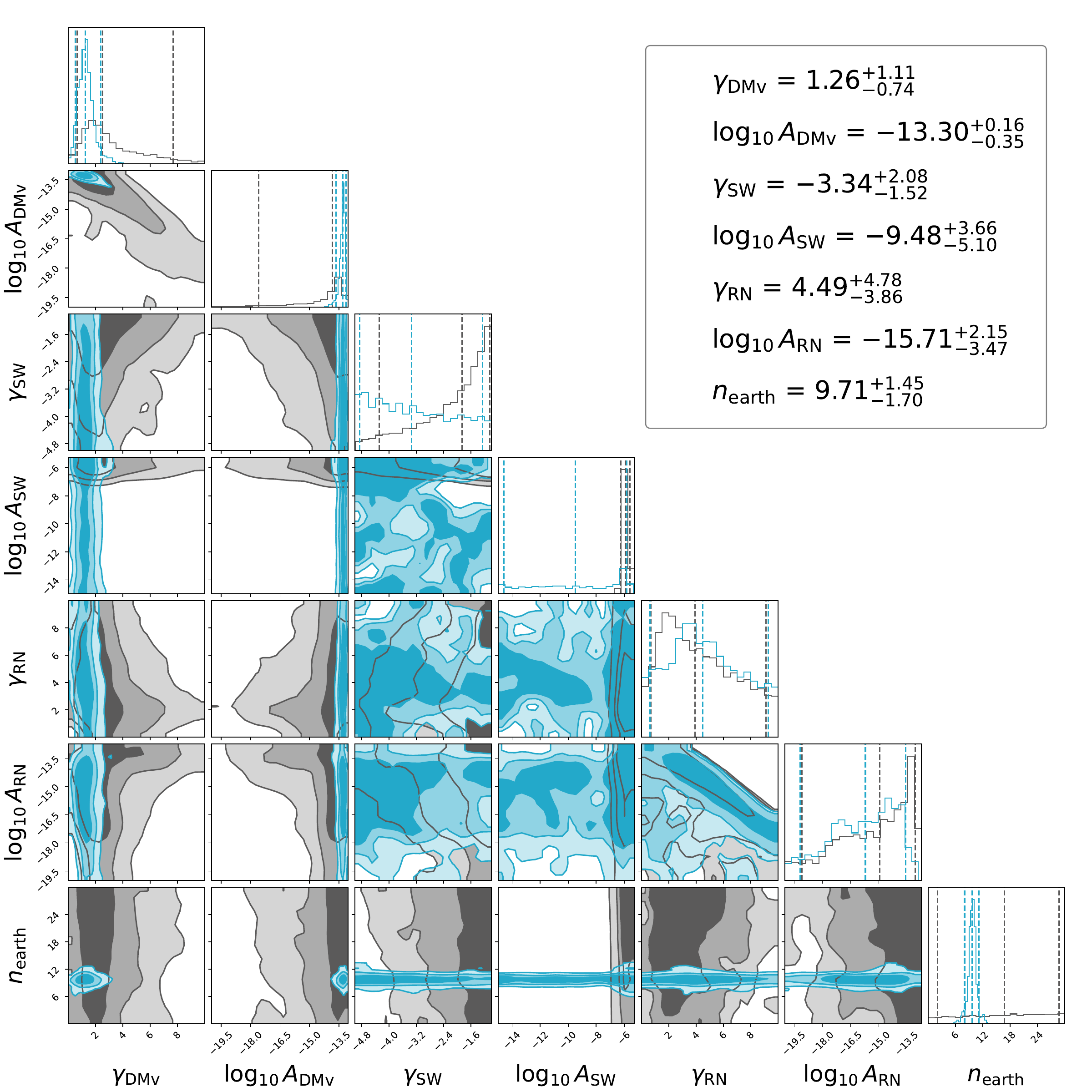}

    \vspace{0.1em}
    {\small (b) PSR~J1022$+$1001}


    \includegraphics[width=0.72\columnwidth]
        {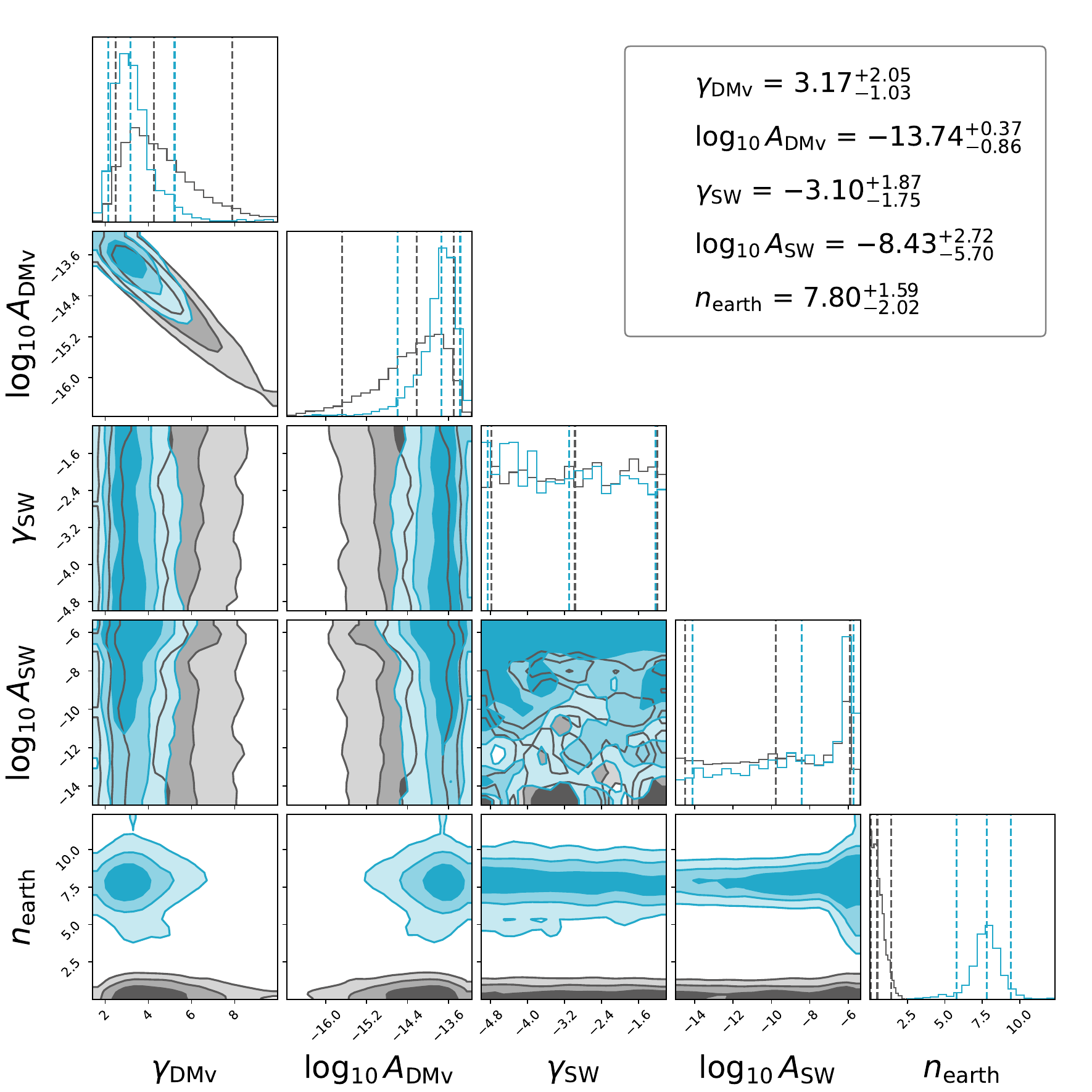}

    \vspace{0.1em}
    {\small (c) PSR~J1730$-$2304}

    \caption{Corner plots of the SP parameters of the
    highest-posterior 2023 noise model, evaluated using the 2020~DR2 (grey) and the 2023~DR2 (teal) pipelines:
    (a)~PSR~J1909$-$3744, representing a decisive model space;
    (b)~PSR~J1022$+$1001, representing a competitive space;
    and (c)~PSR~J1730$-$2304, which transitions from a diffuse to a
    competitive space.}
    \label{fig:corner_plots}
\end{figure}

\subsection{Classification of PPTA DR2 model spaces}

The complete model-posterior distributions in
Figures~\ref{fig:model_probs}, \ref{fig:app_model_1}, and \ref{fig:app_model_2} show that the
changes between pipelines take several distinct forms. For some
pulsars, the same broad stochastic-process family remains preferred
and the posterior concentration changes only moderately. In other
cases, posterior support is redistributed among models that contain
the same stochastic processes but differ in their white-noise
components. A third group exhibits a qualitative change in the
stochastic-process composition, most commonly through increased
support for models containing SW.

Several model spaces remain broad under both pipelines. For
example, the model posteriors of PSRs J1024$-$0719,
J1045$-$4509, and J1446$-$4701 remain distributed among numerous
models rather than becoming dominated by a single configuration. The
absence of a decisive model in such cases is itself an inference from
the data: several combinations of white noise, RN, and DMv remain
plausible within the candidate set.

At the opposite extreme, pulsars such as J1713+0747,
J1744$-$1134, J1909$-$3744, and J1939+2134 retain strongly
concentrated model posteriors. Nevertheless, concentration of the
model posterior does not necessarily imply complete stability under
reprocessing. The identity of the preferred white-noise variant or
the inclusion of a solar-wind component can change even when
$p_{\rm top}$ remains in the decisive range. The three pulsars
examined below illustrate the different forms of stability,
competition, and posterior reorganization visible in the full sample.

\subsubsection{A decisive model space: PSR J1909$-$3744}

PSR J1909$-$3744 provides an example in which the inferred
stochastic-process composition is stable while the preferred
white-noise description changes. As shown in
Figure~\ref{fig:model_probs}(a), an RN+DMv+SW model family
dominates the model space under both pipelines.
The EC+RN+DMv+SW model carries approximately $92\%$ of the posterior
probability in the 2020 analysis, whereas the corresponding model
including EQ carries approximately $91\%$ in the 2023 analysis.

The change in the highest-probability model is therefore associated
primarily with the additional white-noise component rather than with
a change in the inferred set of stochastic processes. This example
demonstrates why it is useful to distinguish between the complete
noise-model label and the underlying stochastic-process family.
At the level of the latter, the model selection is highly stable.

The parameter posteriors in Figure~\ref{fig:corner_plots}(a)
provide additional information that is not visible from the model
probabilities alone. The RN and DMv parameters are constrained under
both pipelines and remain mutually consistent. In contrast, the
solar-wind parameters becomes more informative. In
the 2020 analysis, the solar-wind posterior is comparatively broad
and largely prior dominated. After reprocessing, the posterior
becomes localized, yielding $
 n_{\rm earth}=3.81^{+0.50}_{-0.51}\ {\rm cm}^{-3}.$
Thus, even though the same stochastic-process family dominates both model spaces, the reprocessed observations provide substantially more information about the solar-wind contribution.

\subsubsection{A competitive model space: PSR J1022+1001}

PSR J1022+1001 illustrates a different form of model uncertainty.
The model posterior is concentrated on four SW models in
both datasets Figure~\ref{fig:model_probs}(b). In the 2020 analysis,
EC+RN+DMv+SW and EC+DMv+SW have posterior probabilities of
approximately $0.41$ and $0.34$, respectively. In the 2023 analysis,
the two models become nearly degenerate, each carrying approximately
$0.43$ of the posterior probability. The remaining support is divided between the EQ+EC+RN+DMv+SW and EC+RN+SW models.  Within the candidate model set and under the adopted uniform model priors, the solar-wind contribution is robustly supported even though the complete noise model remains uncertain.  The unresolved question is whether an additional achromatic RN process is required. Model competition therefore occurs mainly between the RN-containing and
RN-free models, rather than between models with and without a DM and SW contributions.

This interpretation is consistent with
Figure~\ref{fig:corner_plots}(b). The DMv parameters and
$n_{\rm earth}$ are well constrained by the reprocessed data, whereas
the RN amplitude remains weakly constrained. Consequently, adding RN
does not lead to a sufficiently distinct description for the data to
select decisively between the two leading models. Conditional on the EC+RN+DMv+SW model, the inferred solar-wind density is  $n_{\rm earth}=9.71^{+1.45}_{-1.70}\ {\rm cm}^{-3}$. This density is consistent with the value obtained from the longer PPTA DR3 dataset \cite{reardon2023gravitational}. PSR J1022+1001 therefore demonstrates that strong support for an individual physical component can coexist with substantial
uncertainty over the complete noise model.
\begin{figure*}
    \centering
    \includegraphics[width=0.8\textwidth]{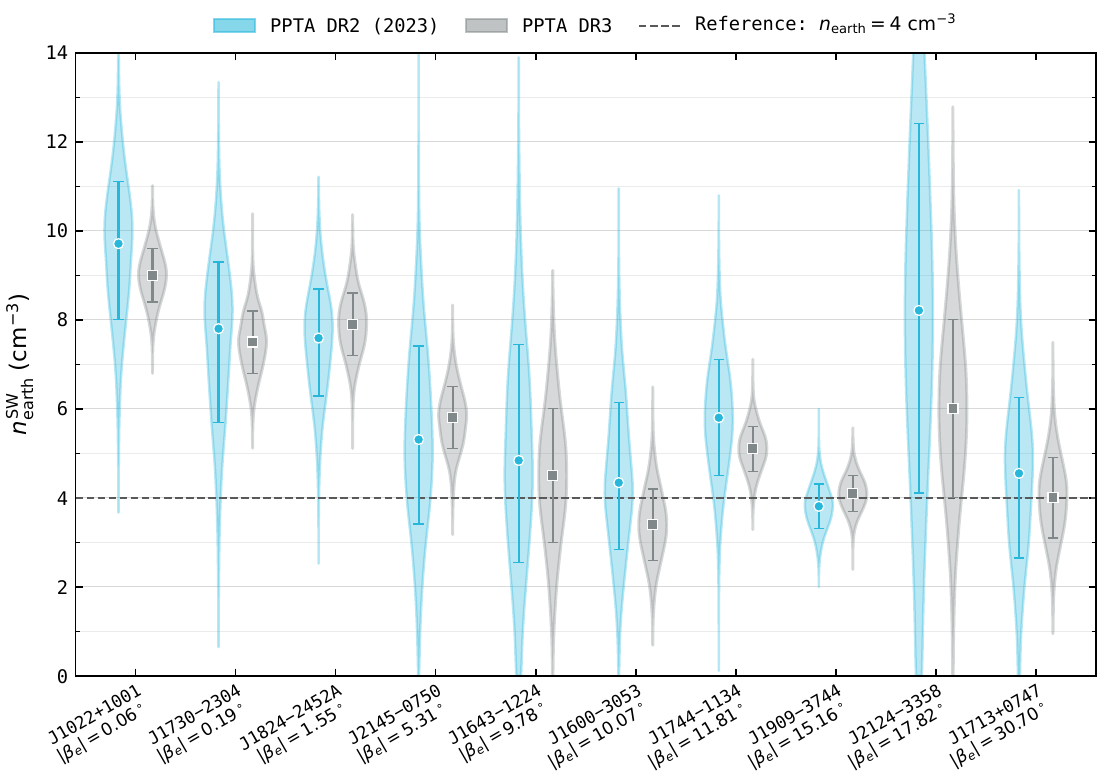}
    \caption{
        Comparison of the inferred solar-wind electron density at Earth,
        $n_{\rm earth}^{\rm SW}$, for ten PPTA pulsars in the PPTA DR2
        (2023) and PPTA DR3 analyses\citep{reardon2023gravitational}. The violin profiles show the
        distributions of $n_{\rm earth}^{\rm SW}$  posterior summaries from Table~\ref{tab:sw_comparison},
        while the markers and vertical bars indicate the posterior medians
        and central $68\%$ credible intervals. The horizontal dashed line
        marks the reference value $n_{\rm earth}=4\,{\rm cm}^{-3}$.
        Pulsars are arranged by increasing absolute ecliptic latitude,
        $|\beta_e|$, which is reported below each pulsar.
    }
    \label{fig:nearth_ppta_dr2_dr3_comparison}
\end{figure*}
\subsubsection{Posterior reorganization: PSR J1730$-$2304}

PSR J1730$-$2304 illustrates how reprocessing can reorganize a
previously diffuse model posterior. Under the 2020 pipeline, the
posterior probability is divided among several DMv-containing models,
with no model reaching the competitive threshold. The hypothesis
space therefore does not provide a clear preference for a particular
combination of DMv, RN, and white-noise components.

In the 2023 analysis, the support becomes concentrated on models
containing both DMv and SW. The EC+DMv+SW model carries approximately
$60\%$ of the posterior probability, causing the pulsar to move from
the diffuse to the competitive class. The posterior does not become
fully decisive, because non-negligible probability remains assigned
to alternative SW-containing configurations. Nevertheless, the
reprocessed data identify a much more restricted region of the
model space than the original pipeline.

The parameter posterior in Figure~\ref{fig:corner_plots}(c) mirrors
this model-level change. When both data sets are evaluated under the highest-posterior 2023 model, the solar-wind-density posterior is broad under the 2020 pipeline but becomes substantially more localized under the 2023 pipeline, with $
 n_{\rm earth}=7.80^{+1.59}_{-2.02}\ {\rm cm}^{-3}.$
This is consistent with the PPTA DR3 \cite{reardon2023gravitational} estimate of
$7.5\pm0.7\ {\rm cm}^{-3}$. The agreement indicates that the
solar-wind information recovered in the longer DR3 analysis is
already present in the reprocessed DR2 observations.

\subsection{Solar-wind emergence in PPTA DR2}

Table~\ref{tab:sw_comparison} compares the 2023 DR2 estimates of
$n_{\rm earth}$ with those obtained from the full PPTA DR3 dataset \citep{reardon2023gravitational}.
The table contains ten pulsars spanning absolute ecliptic latitudes
from $0.06^\circ$ to $30.70^\circ$. For every pulsar, the $n_{\rm earth}$ posterior
median inferred from the reprocessed DR2 data is consistent with the
corresponding DR3 estimate within the reported credible intervals.

Figure~\ref{fig:nearth_ppta_dr2_dr3_comparison}  displays the solar wind agreement between datasets, despite the shorter observing baseline of
the DR2 subset and the absence of the UWL observations included in DR3 \citep{reardon2023gravitational}. The DR2 credible intervals are generally wider, as expected from
the reduced data volume, but they are centred on values compatible
with the ones from the longer dataset. The comparison therefore shows that the
increased support for SW models is not produced solely by the
additional years of DR3 observations. Instead, improved processing
of the observations already contained in DR2 is sufficient to reveal
the solar-wind contribution.

The posterior distributions for the seven additional pulsars in
Figure~\ref{fig:app_corner_sw} support the same interpretation.
Together with the three representative pulsars in
Figure~\ref{fig:corner_plots}, they show that the 2023 analysis
generally yields more localized solar-wind-density posteriors than
the original pipeline. The improvement is not identical for every
pulsar, and the widths of the distributions continue to depend on
the available information for each line of sight. Nevertheless, the
agreement with the independently obtained DR3 values across all ten
pulsars establishes the consistency of the recovered signal.

\begin{table}
\centering
\caption{Solar-wind electron density at 1\,au, $n_{\mathrm{earth}}$ ($\mathrm{cm}^{-3}$), $n_{\mathrm{earth}}^{\mathrm{DR2}}$ from the highest-posterior SW-containing model under the 2023 DR2 analysis and $n_{\mathrm{earth}}^{\mathrm{DR3}}$  the PPTA~DR3 analysis \citep{reardon2023gravitational}.
Posterior medians and central 68\% credible intervals are shown; pulsars are sorted by absolute ecliptic latitude $|\beta_e|$ in degrees.}
\label{tab:sw_comparison}
\renewcommand{\arraystretch}{1.5}
\begin{tabular}{lccc}\toprule

Pulsar & $|\beta_e|$& $n_{\mathrm{earth}}^{\mathrm{DR2}}$ & $n_{\mathrm{earth}}^{\mathrm{DR3}}$ \\\midrule

J1022$+$1001  & $0.06$  & $9.71^{+1.4}_{-1.7}$& $9.0^{+0.6}_{-0.6}$ \\
J1730$-$2304  & $0.19$  & $7.80^{+1.5}_{-2.1}$& $7.5^{+0.7}_{-0.7}$ \\
J1824$-$2452A & $1.55$  & $7.59^{+1.1}_{-1.3}$& $7.9^{+0.7}_{-0.7}$ \\
J2145$-$0750  & $5.31$  & $5.31^{+2.1}_{-1.9}$& $5.8^{+0.7}_{-0.7}$ \\
J1643$-$1224  & $9.78$  & $4.84^{+2.6}_{-2.3}$& $4.5^{+1.5}_{-1.5}$ \\
J1600$-$3053  & $10.07$ & $4.34^{+1.8}_{-1.6}$& $3.4^{+0.8}_{-0.8}$ \\
J1744$-$1134  & $11.81$ & $5.80^{+1.3}_{-1.3}$& $5.1^{+0.5}_{-0.5}$ \\
J1909$-$3744  & $15.16$ & $3.81^{+0.5}_{-0.5}$& $4.1^{+0.4}_{-0.4}$ \\
J2124$-$3358  & $17.82$ & $8.21^{+4.2}_{-4.7}$& $6.0^{+2.0}_{-2.0}$ \\
J1713$+$0747  & $30.70$ & $4.55^{+1.7}_{-1.9}$& $4.0^{+0.9}_{-0.9}$ \\ \bottomrule
\end{tabular}
\end{table}
\section{Conclusion}
\label{sec:conclusion}

In this work, we have examined how the processing of a pulsar-timing dataset affects Bayesian noise-model inference by comparing the model spaces obtained from two processing versions of the PPTA second data release. For each of 22 pulsars, we evaluated the same set of 13 candidate noise models and used independently estimated marginal
likelihoods to construct the complete model posterior.  The comparison demonstrates that preprocessing choices can materially
alter both the concentration and the composition of the inferred
noise-model posterior. Seven pulsars change between the decisive,
competitive, and diffuse classes. These transitions occur in both
directions: for some pulsars, reprocessing concentrates support on a
smaller set of models, whereas for others it distributes the support
more broadly. Improved processing therefore does not merely increase
the apparent decisiveness of model selection. Instead, it changes
which distinctions between the candidate stochastic processes are
supported by the observations.

The clearest systematic change is the increased posterior support of SW-containing models.  A solar-wind component appears in the
highest-posterior model for ten pulsars in the 2023 analysis, compared
with four under the original pipeline. The inferred values of the
mean electron density at 1 AU are consistent with the values reported
from the longer PPTA DR3 dataset for all ten pulsars \cite{reardon2023gravitational}. Although the
credible intervals from the shorter DR2 baseline are generally wider,
their agreement with the DR3 measurements shows that the solar-wind
signal is already present in the DR2 observations.

The results also illustrate why support for a physical noise component
should not be equated with decisive selection of a complete model.
For J1022+1001, for example, nearly degenerate models provide strong
and stable support for SW while differing over the inclusion of
achromatic RN. Conversely, for J1909$-$3744, the stochastic-process
family remains stable even though the preferred white-noise extension
changes. The full model posterior separates these different forms of
uncertainty, whereas a comparison based only on the
highest-probability model would obscure them.

Most pulsars in both analyses have competitive or diffuse hypothesis
spaces. Averaging over the two pipelines, approximately $75\%$ of the
pulsars are not adequately summarized by a single overwhelmingly
preferred noise model. Selecting only the highest-probability model
for these pulsars would overstate confidence in a particular noise
description and would fail to propagate the uncertainty associated
with alternative, data-supported models.

This issue is directly relevant to PTA searches for a common
red-spectrum process and a gravitational-wave background. Noise
components inferred for individual pulsars can be covariant with a common
process, and a noise model selected in a preliminary analysis is
usually treated as fixed in the subsequent PTA-wide analysis.
Uncertainty that is removed at the single-pulsar stage is therefore
not automatically recovered later. The single-pulsar noise analysis
should instead produce a posterior distribution over plausible noise
models that can be marginalized over, or otherwise propagated, in the
common-process analysis.

The numerical model probabilities reported here are conditional on
the candidate model space and on the uniform model priors adopted
in this work. The decisive, competitive, and diffuse labels are
similarly intended as descriptive categories rather than universal
model-selection thresholds. Because the marginal likelihood of each
model has been estimated independently, the model posterior can be
recomputed under alternative model-prior choices without repeating
the parameter-inference calculations. This provides a practical route
for examining the sensitivity of downstream conclusions to different
physical expectations about the occurrence of RN, DMv, SW, and
additional white-noise components.

More broadly, this study shows that data processing is a component of
the systematic uncertainty budget of Bayesian PTA inference.
Hypothesis-space analysis provides a direct means of exposing that
uncertainty by retaining the relative support for every candidate
model, identifying degeneracies between noise components, and
distinguishing robust component-level conclusions from uncertain
complete-model assignments. As described in \citeauthor{van_haasteren2025} \cite{van_haasteren2025}, propagating this information through
Bayesian model averaging offers a more principled foundation for
gravitational-wave inference than conditioning on a single selected
noise model.

\begin{acknowledgments}
EMZ, PMR, WS, and RM gratefully acknowledge support by the Marsden Fund Council grant MFP-UOA2131 from New Zealand Government funding, managed by the Royal Society Te Apārangi.
We thank the Parkes Pulsar Timing Array collaboration for making their data publicly accessible. This work uses the third data release of the Parkes Pulsar Timing Array \citep{zic2023parkes}, available on the CSIRO Data Access Portal \citep{PPTA_DR3_part2}. We also thank Boris Goncharov for
making the PPTA DR2 noise analysis publicly reproducible
(\url{https://github.com/bvgoncharov/ppta_dr2_noise_analysis}).
This work was performed on the OzSTAR national facility at Swinburne University of Technology. The OzSTAR program receives funding in part from the Astronomy National Collaborative Research Infrastructure Strategy (NCRIS) allocation provided by the Australian Government, and from the Victorian Higher Education State Investment Fund (VHESIF) provided by the Victorian Government.\\

\textit{Software}: \texttt{ENTERPRISE} \cite{ellis2019enterprise,enterprise}, \texttt{enterprise\_extension} \cite{enterprise,justin_ellis_2017_1037579,dalcin2005mpi}, \texttt{Enterprise Warp}, \texttt{EMCEE}\cite{emcee}, \texttt{PocoMC}\cite{karamanis2022pocomc}, \texttt{MorphZ}\cite{morphz}, \texttt{Jupyter} \cite{kluyver2016jupyter}, \texttt{matplotlib} \cite{hunter2007matplotlib}, \texttt{numpy} \cite{harris2020array}, \texttt{scipy} \cite{pauli_virtanen_2020_4406806}, and \texttt{arviz} \cite{kumar2019arviz}.
\end{acknowledgments}
\appendix

\section{Posterior model-probability comparisons}
\label{app:model_probabilities}

\begin{figure*}[p]
  \centering

  \pulsarpanel
    {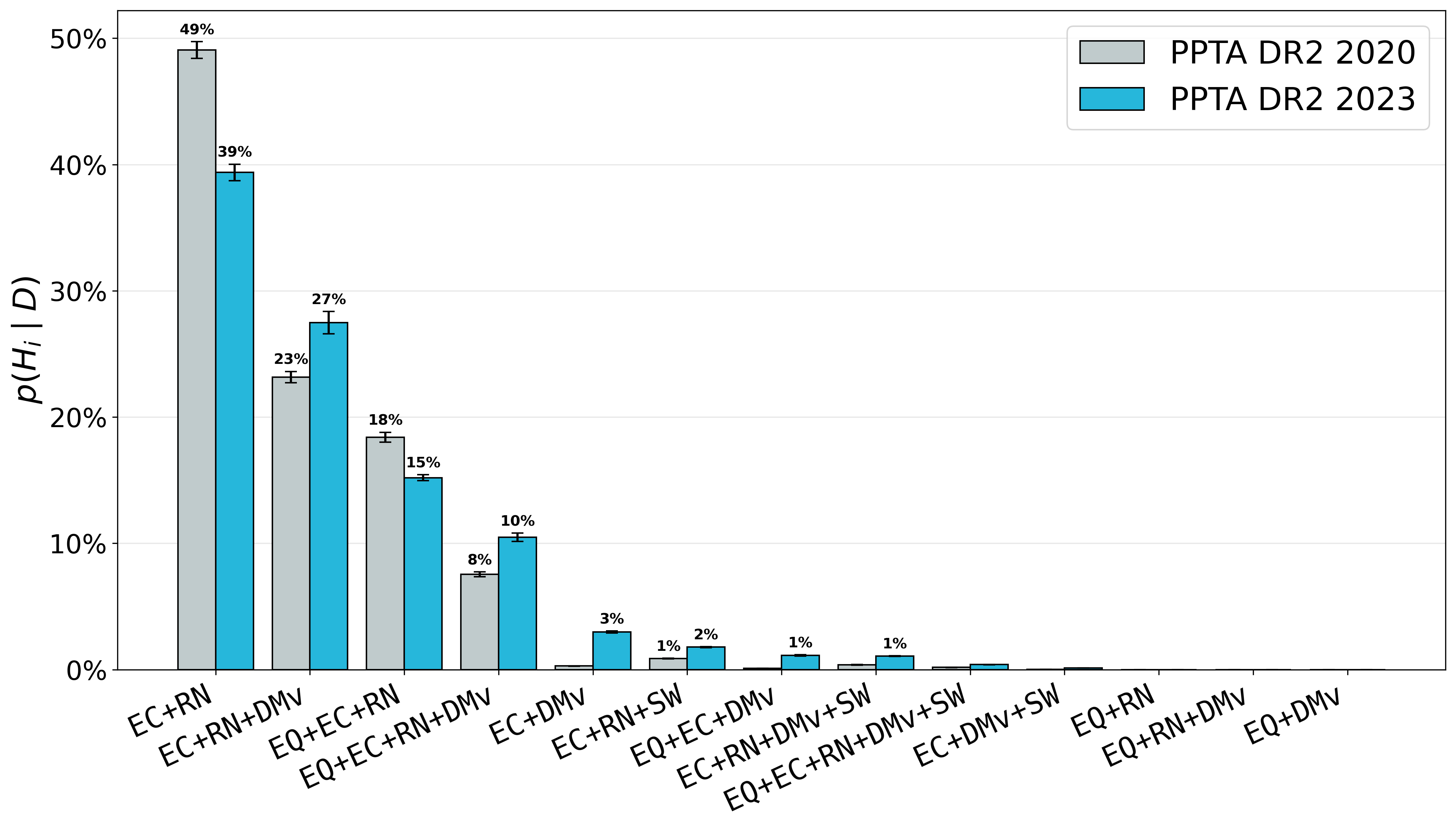}
    {PSR~J0613$-$0200}
  \hfill
  \pulsarpanel
    {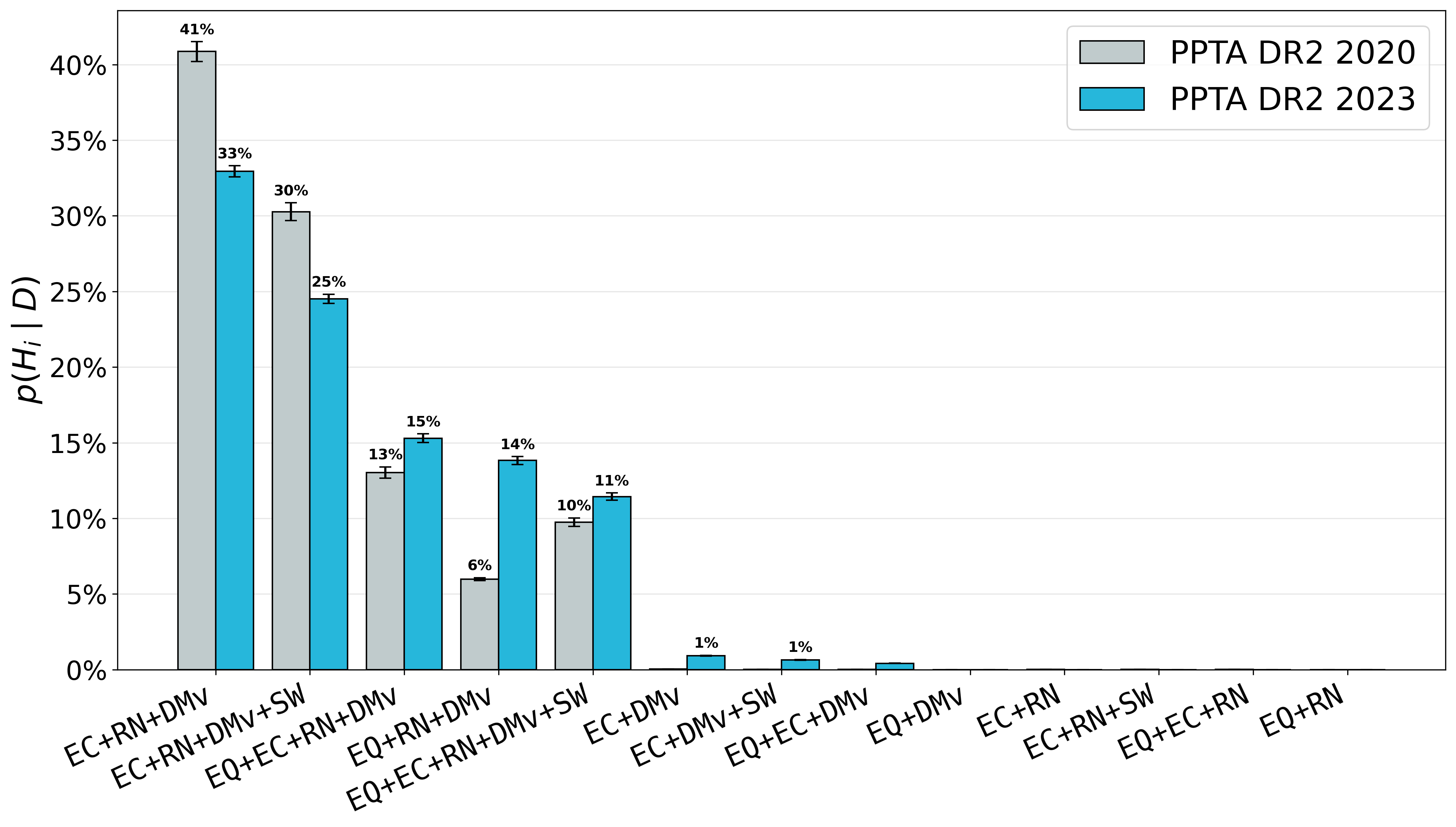}
    {PSR~J0711$-$6830}
  \hfill
  \pulsarpanel
    {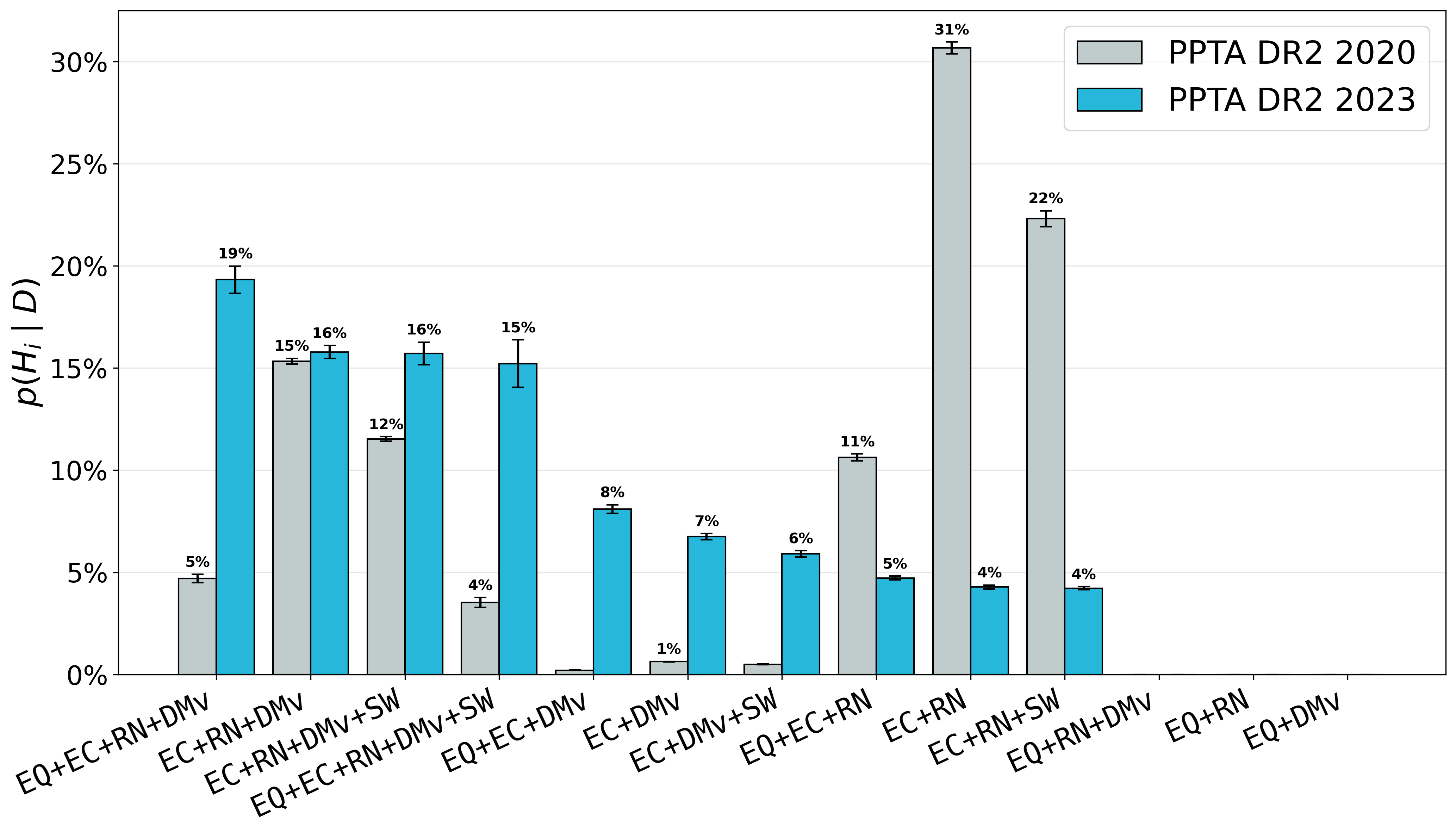}
    {PSR~J1017$-$7156}

  \par\vspace{-1pt}

  \pulsarpanel
    {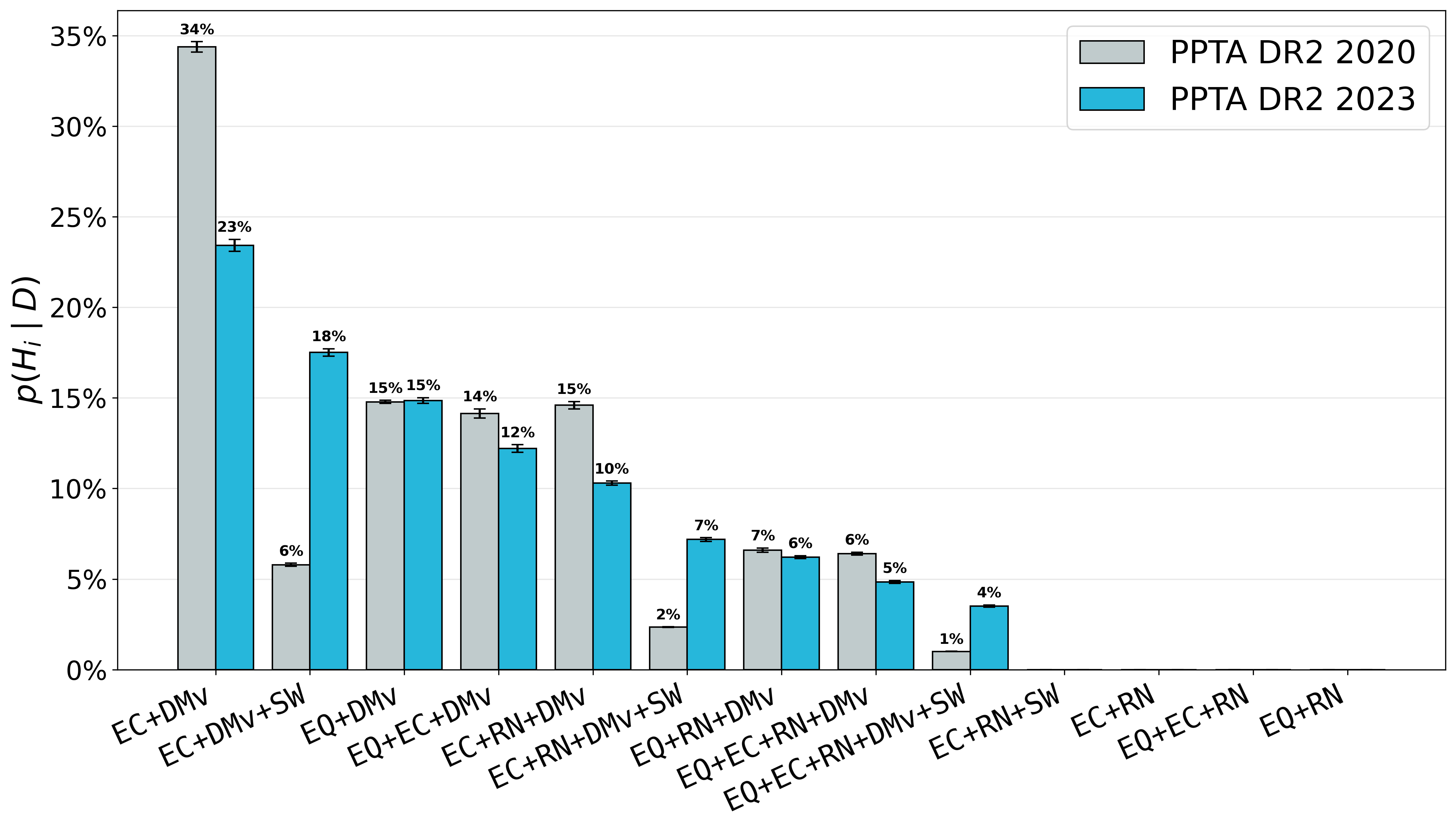}
    {PSR~J1024$-$0719}
  \hfill
  \pulsarpanel
    {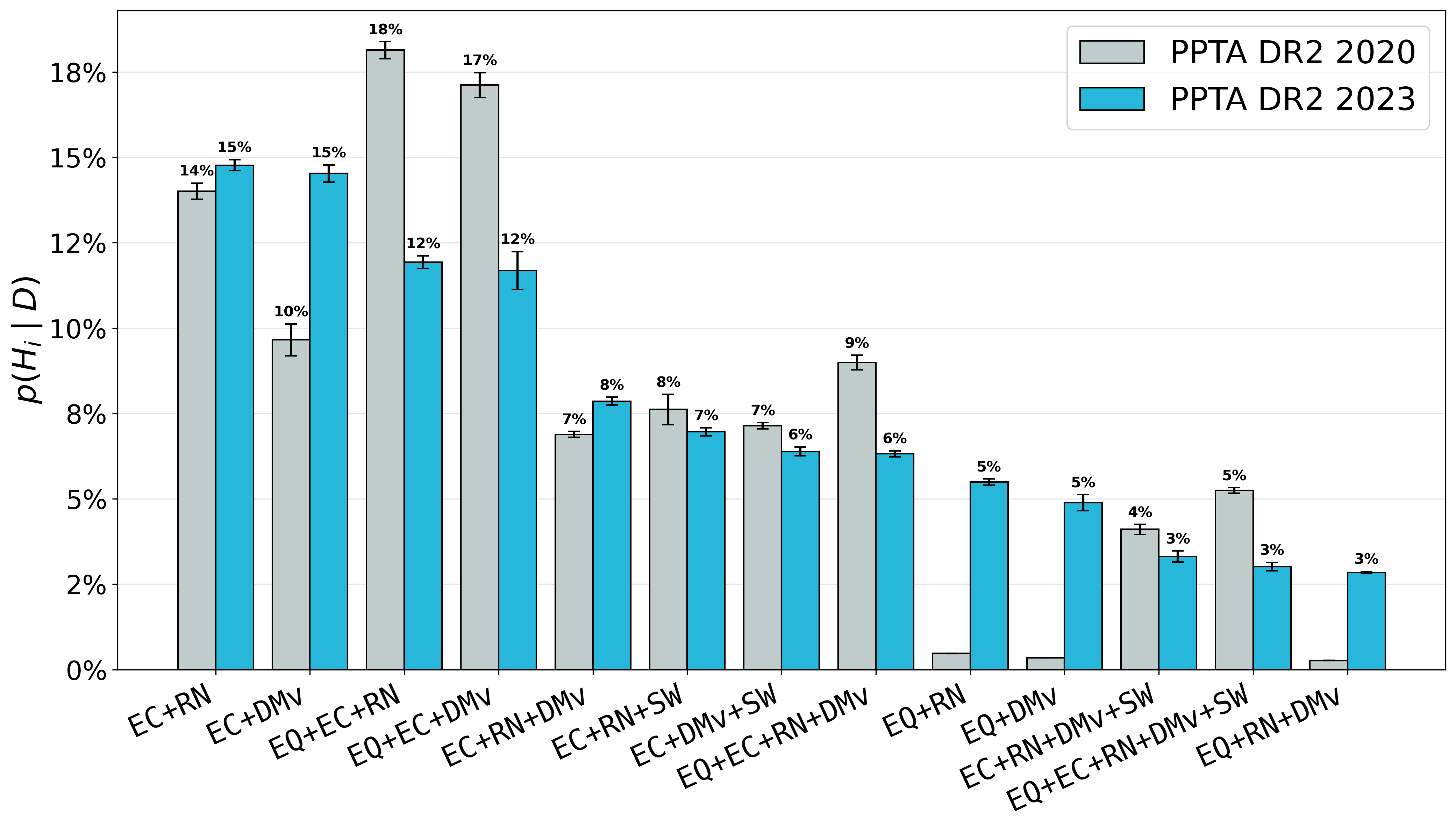}
    {PSR~J1045$-$4509}
  \hfill
  \pulsarpanel
    {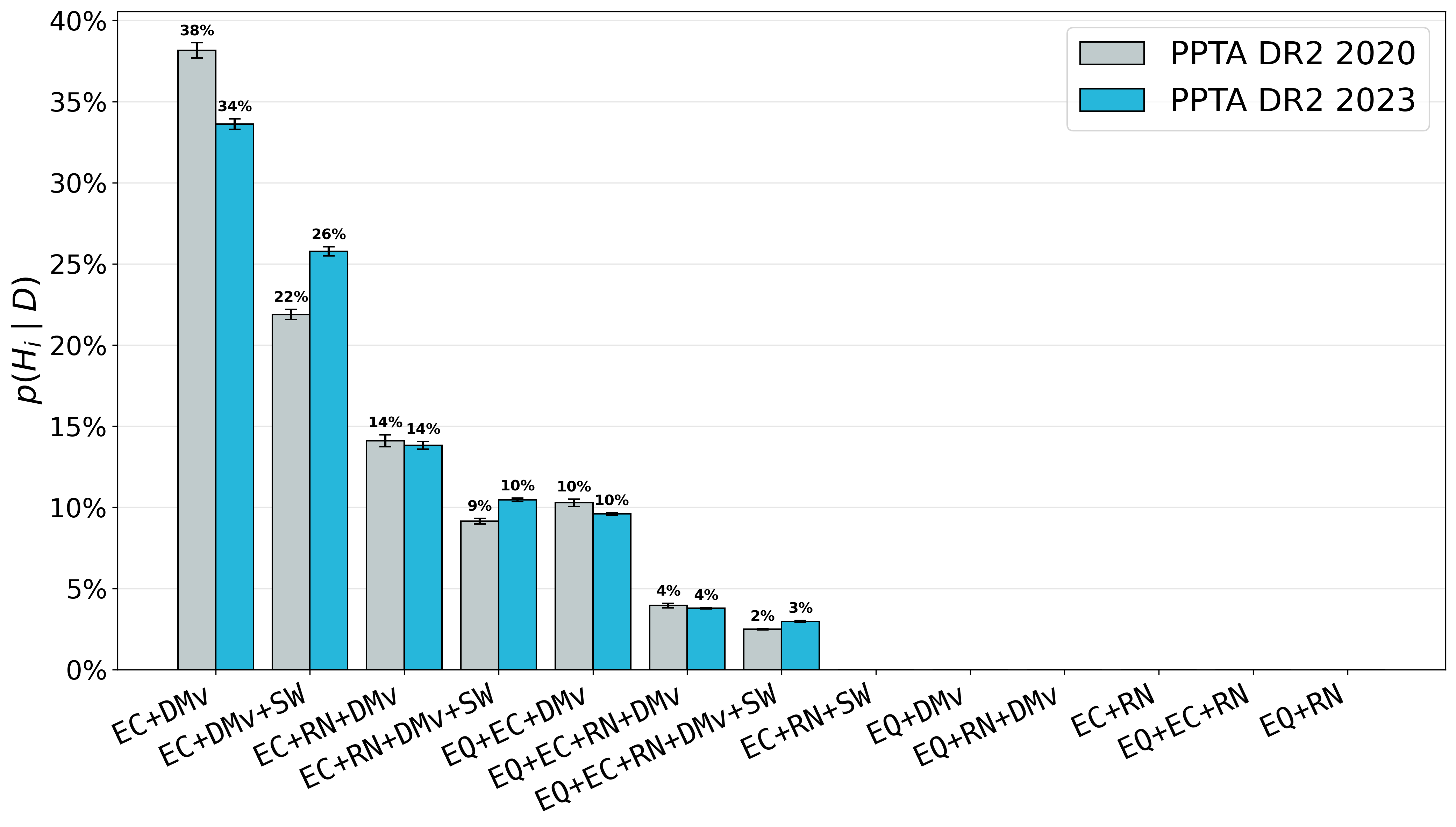}
    {PSR~J1125$-$6014}

  \par\vspace{-1pt}

  \pulsarpanel
    {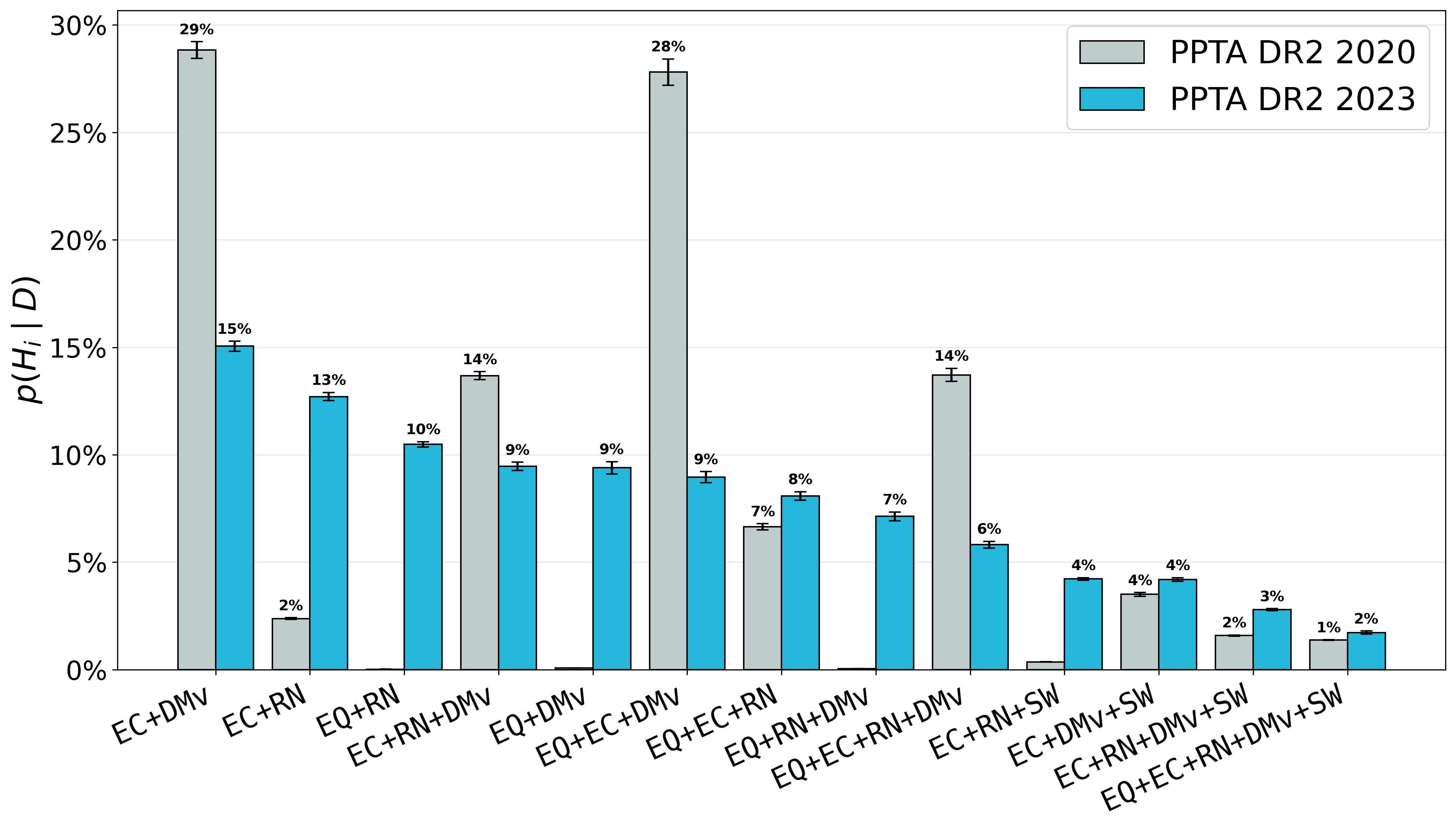}
    {PSR~J1446$-$4701}
  \hfill
  \pulsarpanel
    {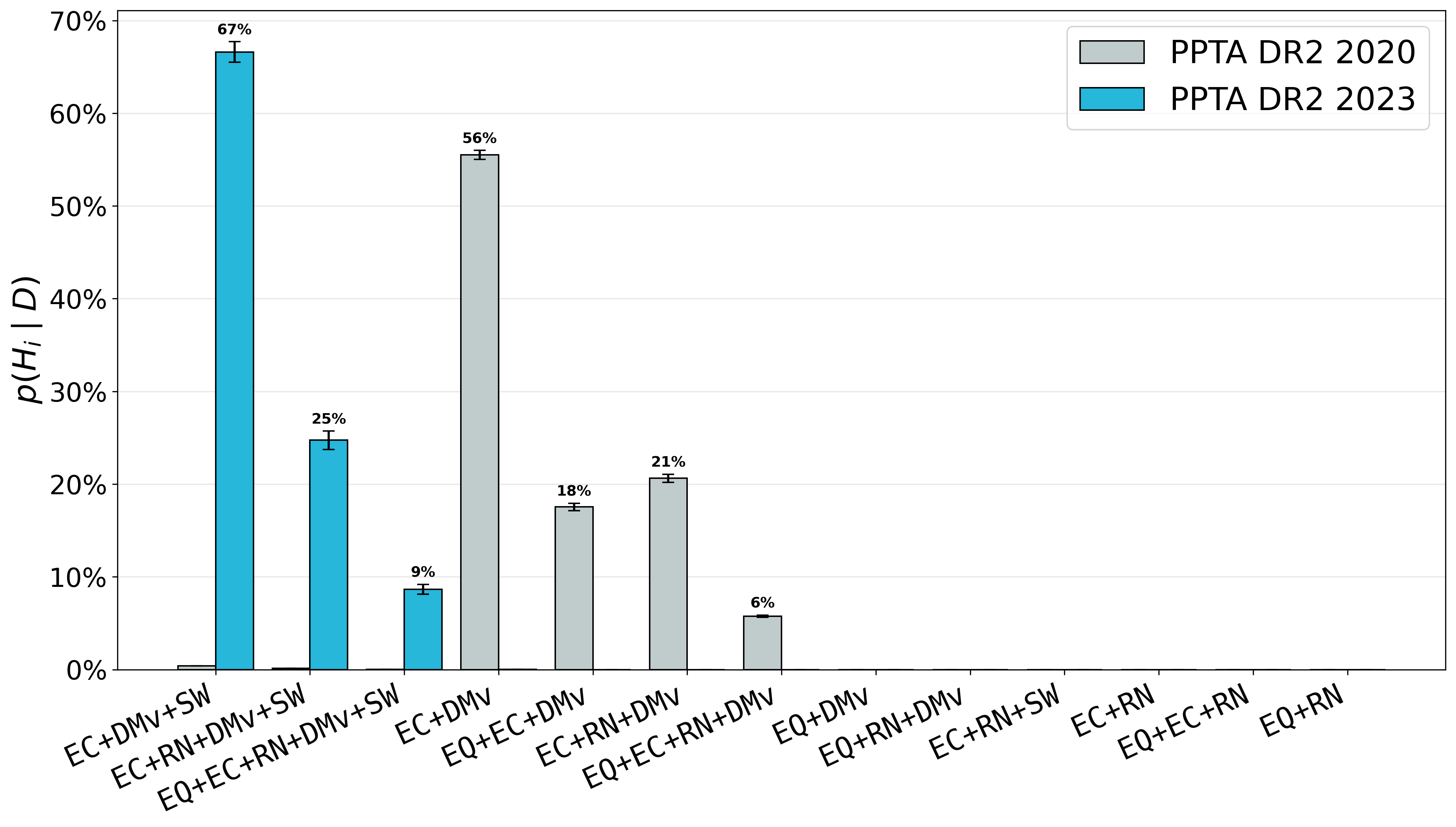}
    {PSR~J1600$-$3053}
  \hfill
  \pulsarpanel
    {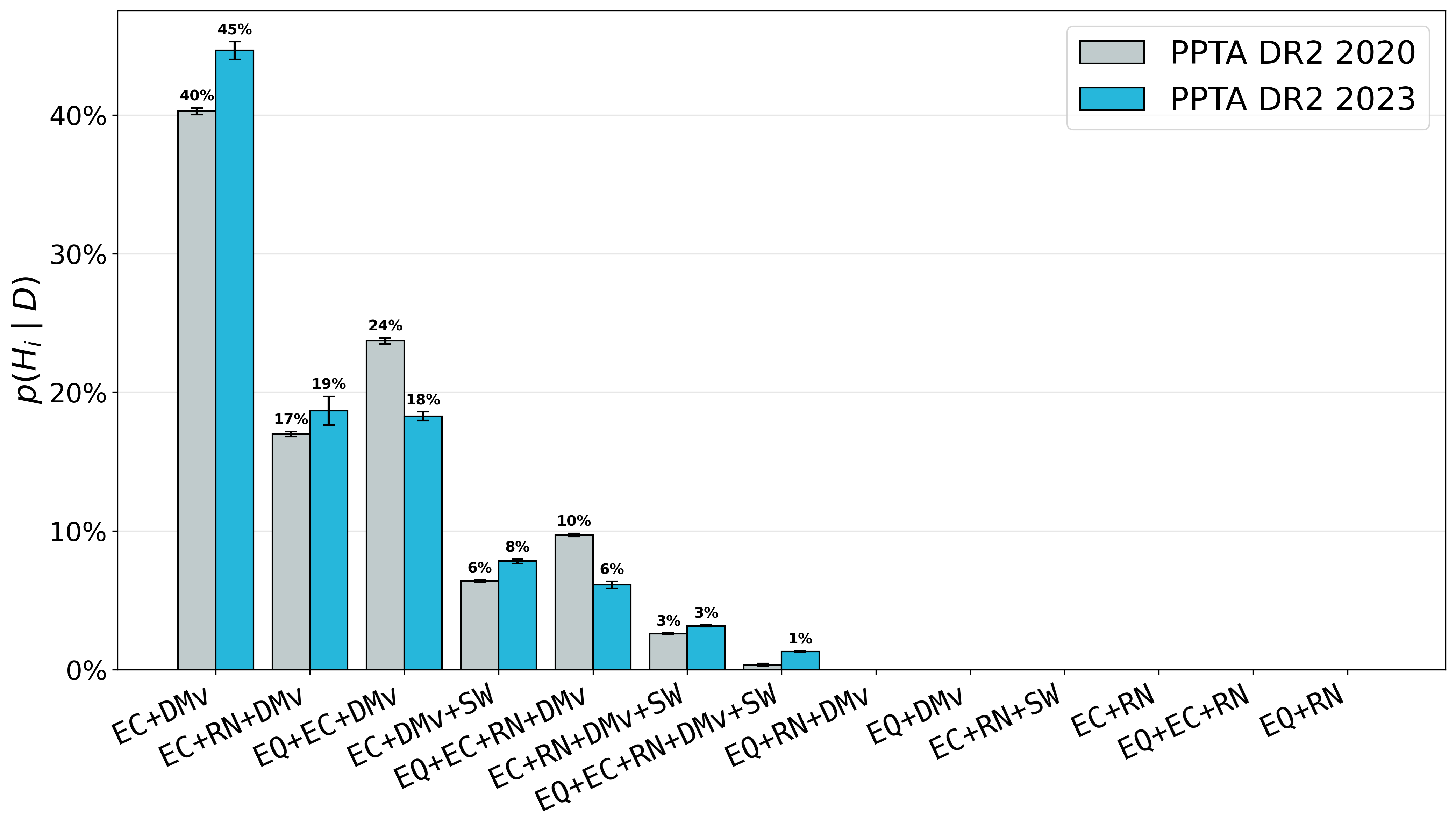}
    {PSR~J1603$-$7202}

  \par\vspace{-1pt}

  \pulsarpanel
    {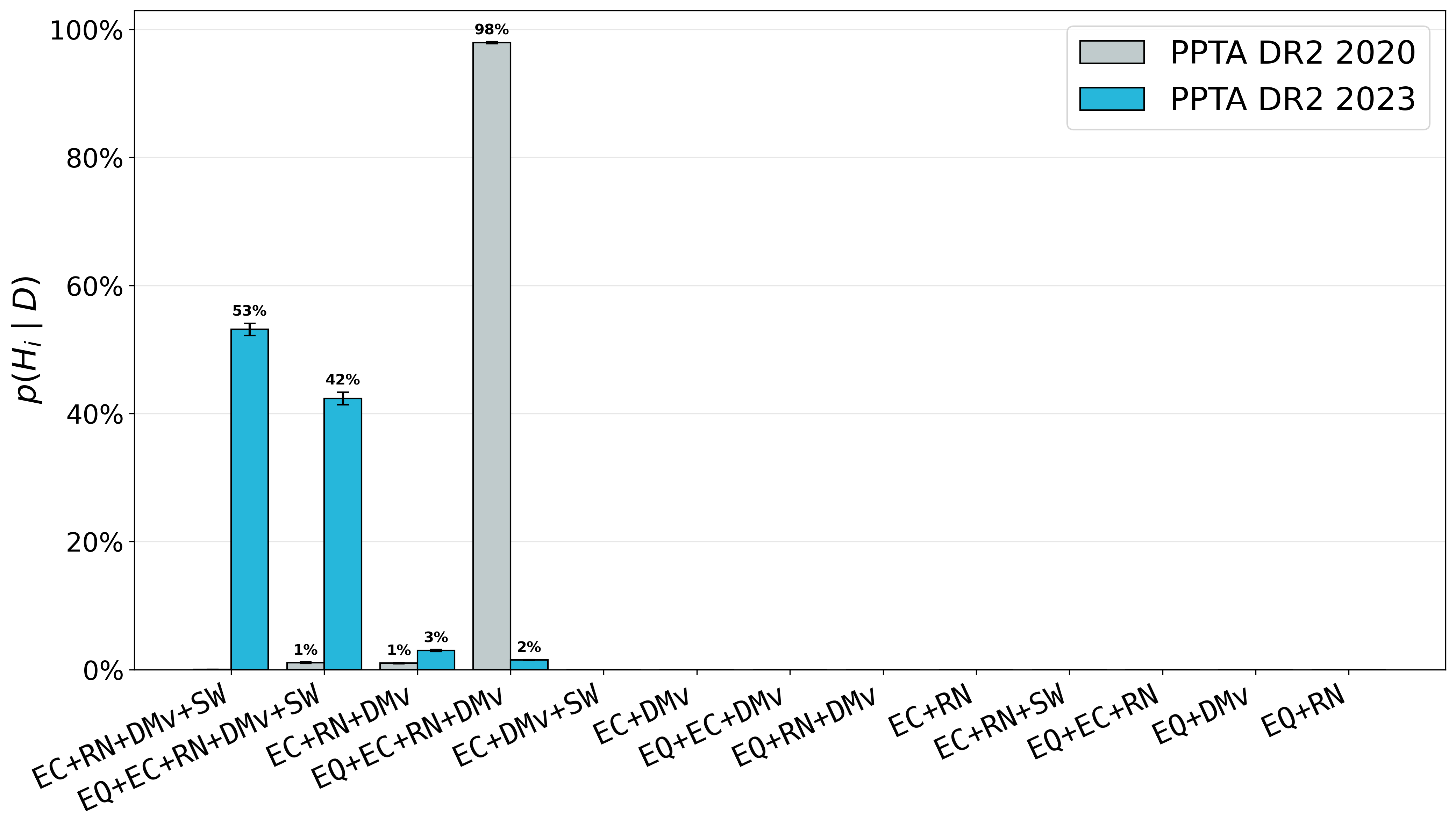}
    {PSR~J1643$-$1224}
  \hfill
  \pulsarpanel
    {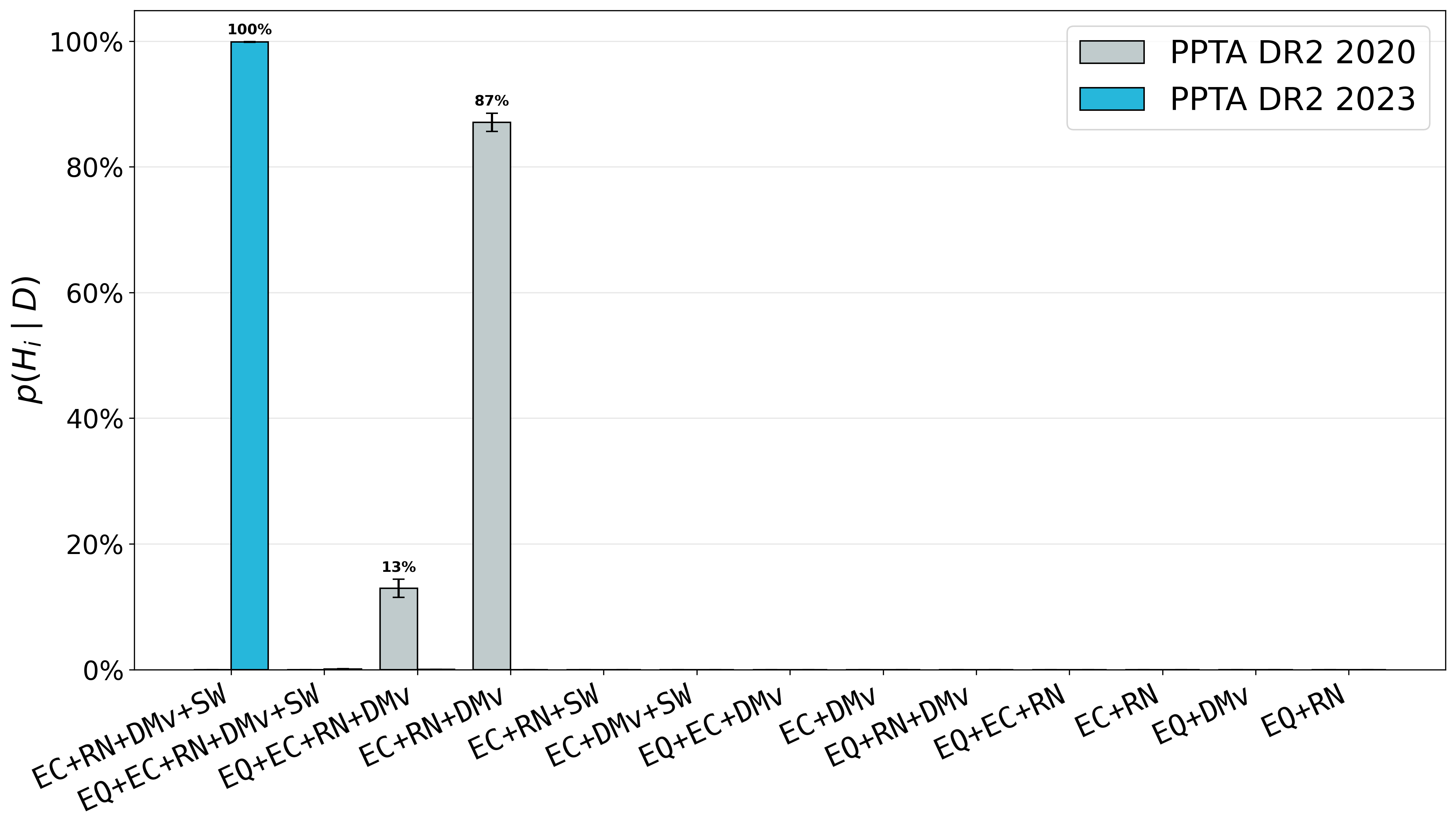}
    {PSR~J1713$+$0747}
  \hfill
  \pulsarpanel
    {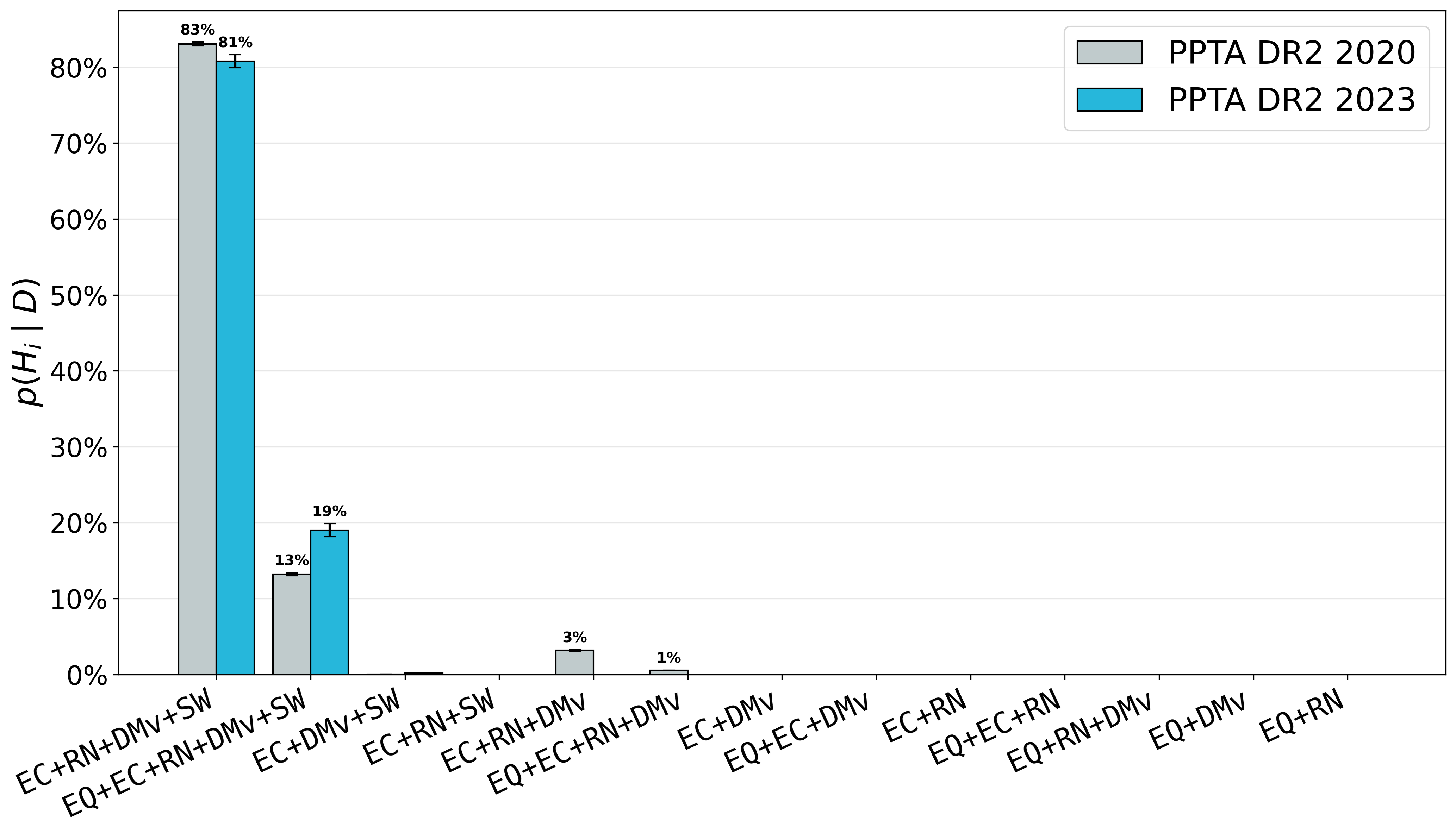}
    {PSR~J1744$-$1134}

  \caption{
    Posterior model probabilities obtained with the 2020~DR2 pipeline
    (grey) and the 2023~DR2 pipeline (teal) for pulsars not shown in
    the main text. This Figure shows the first 12 pulsars.
  }
  \label{fig:app_model_1}
\end{figure*}

\begin{figure*}[p]
  \centering

  \pulsarpanel
    {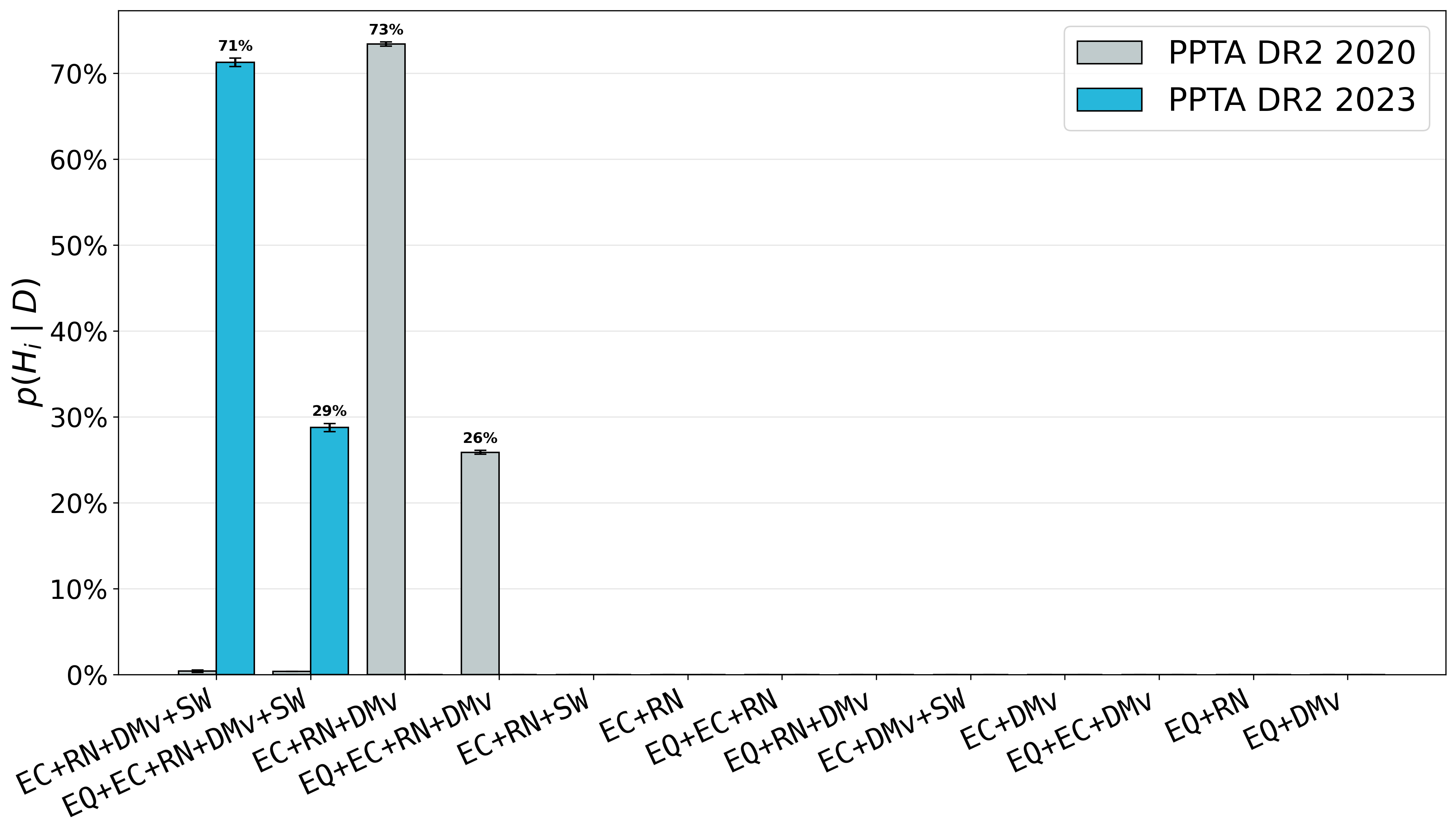}
    {PSR~J1824$-$2452A}
  \hfill
  \pulsarpanel
    {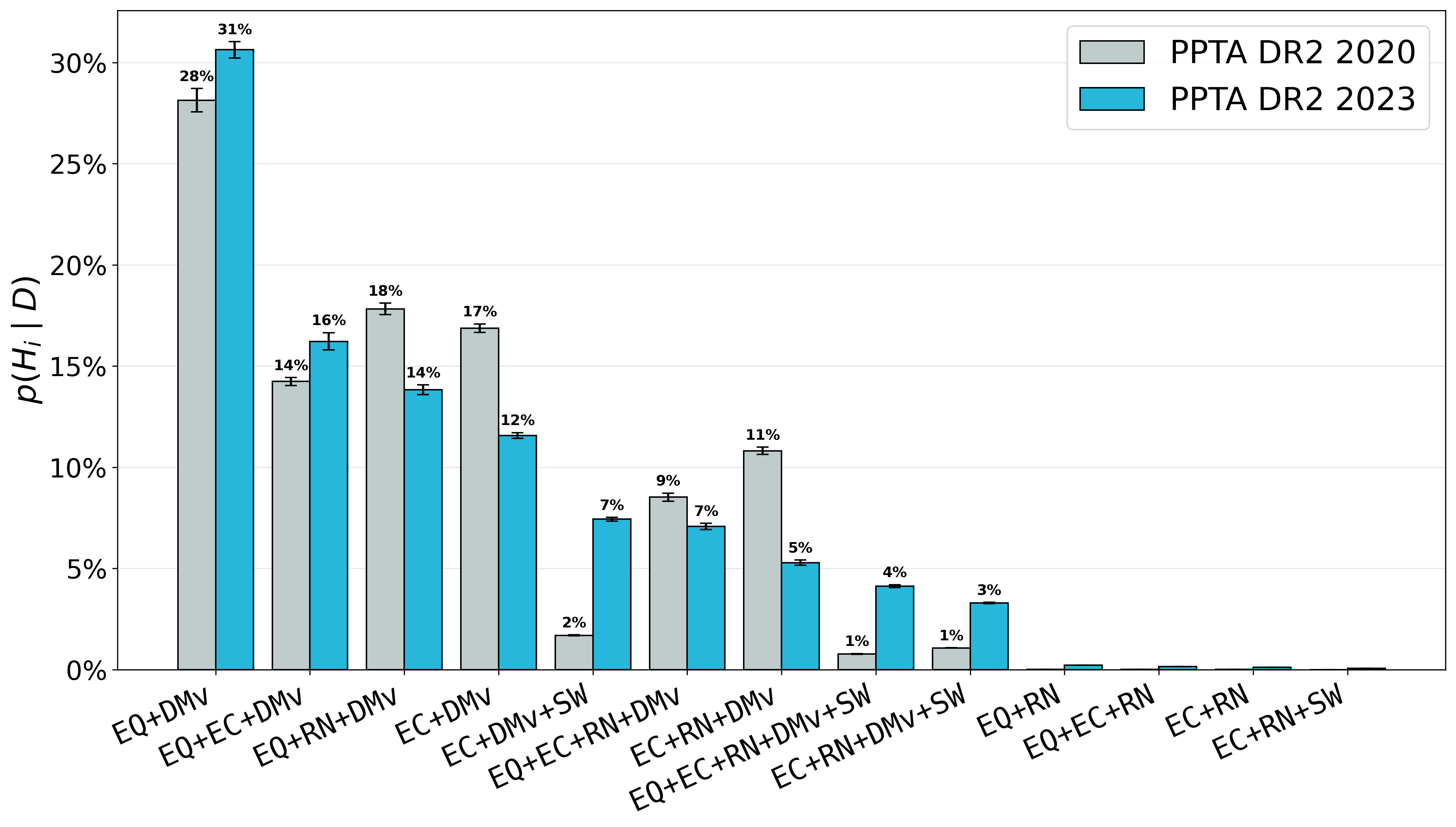}
    {PSR~J1832$-$0836}
  \hfill
  \pulsarpanel
    {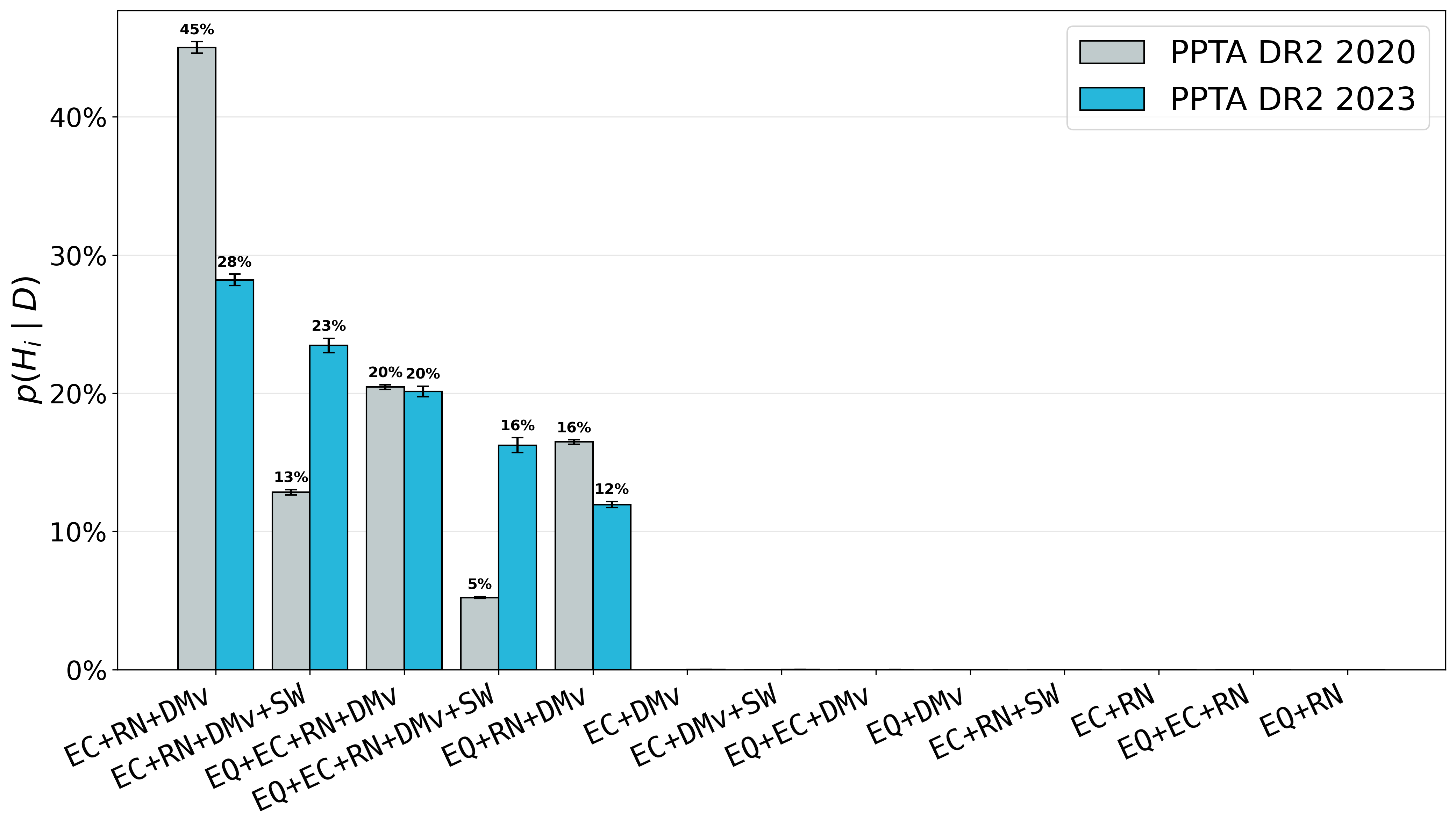}
    {PSR~J1857$+$0943}

  \par\vspace{-1pt}

  \pulsarpanel
    {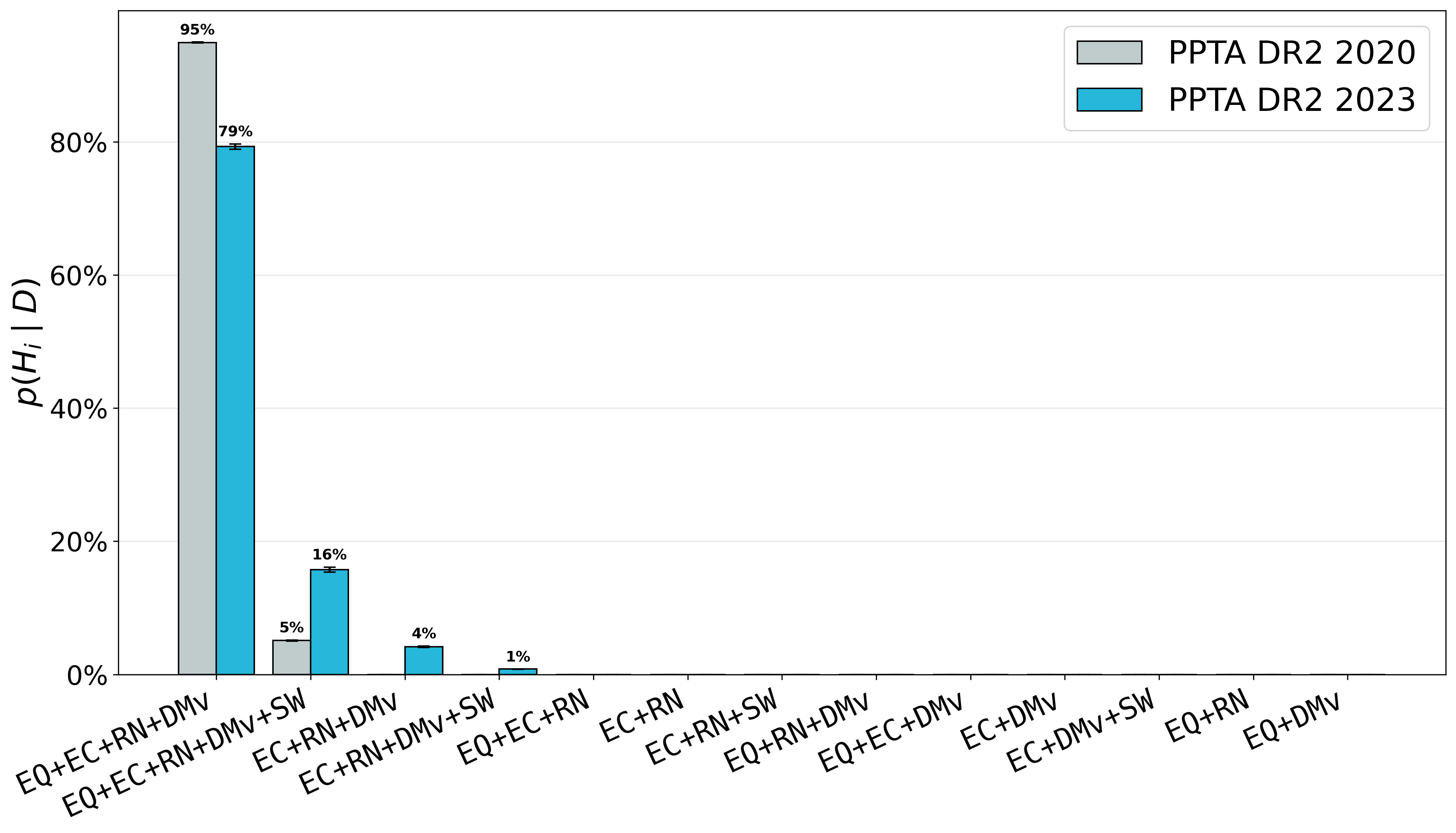}
    {PSR~J1939$+$2134}
  \hfill
  \pulsarpanel
    {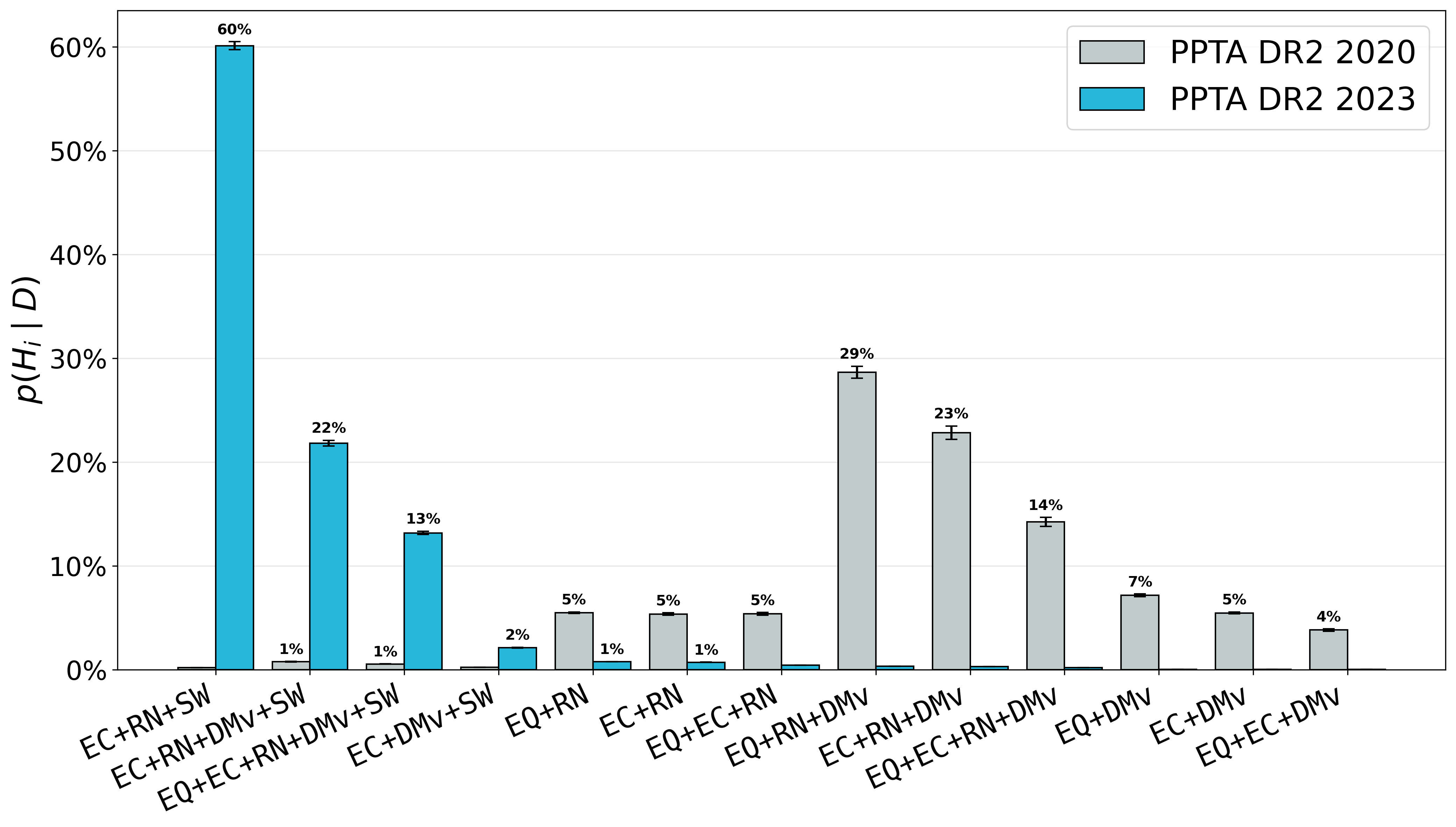}
    {PSR~J2124$-$3358}
  \hfill
  \pulsarpanel
    {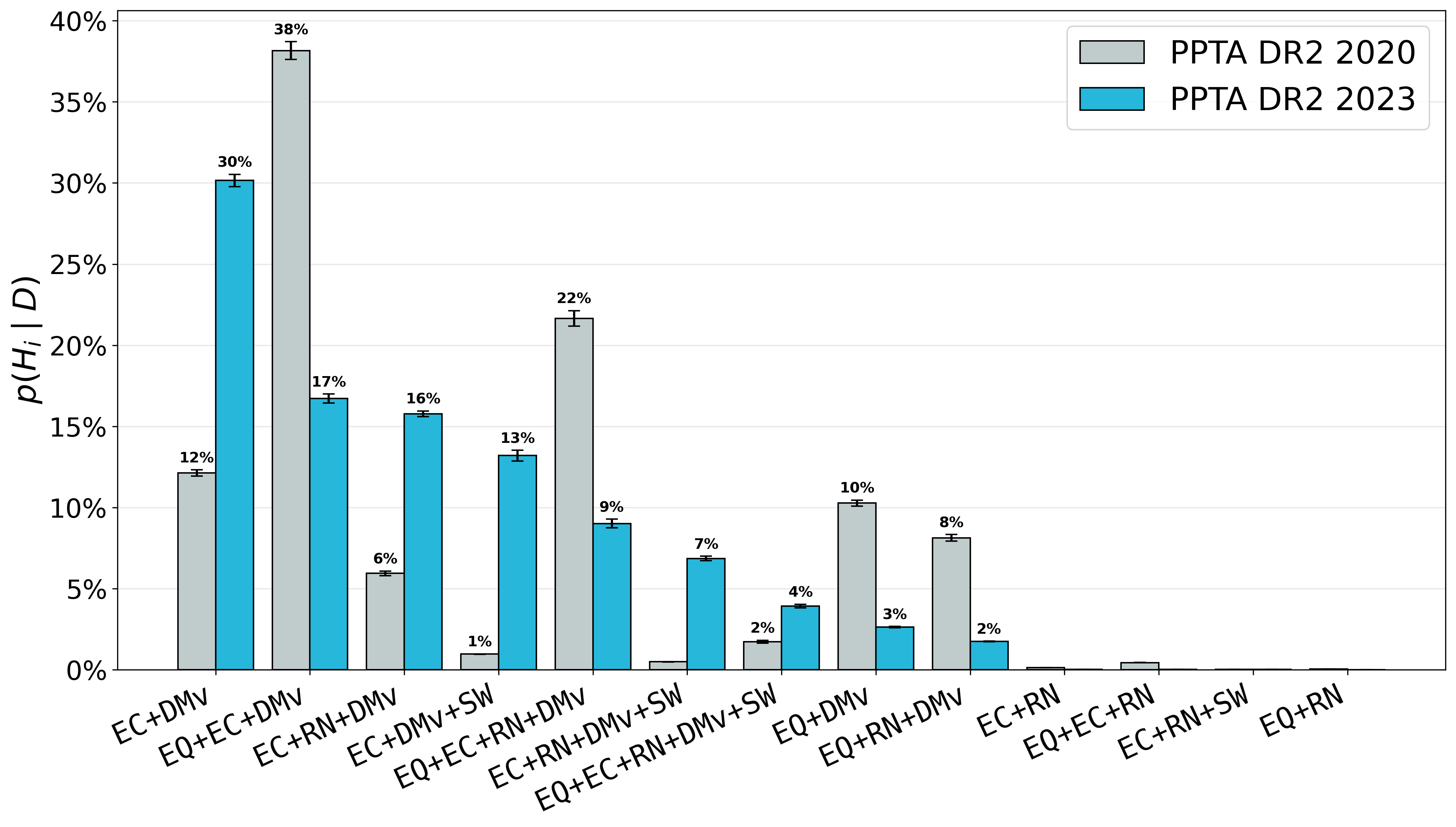}
    {PSR~J2129$-$5721}

  \par\vspace{-1pt}

  \pulsarpanel
    {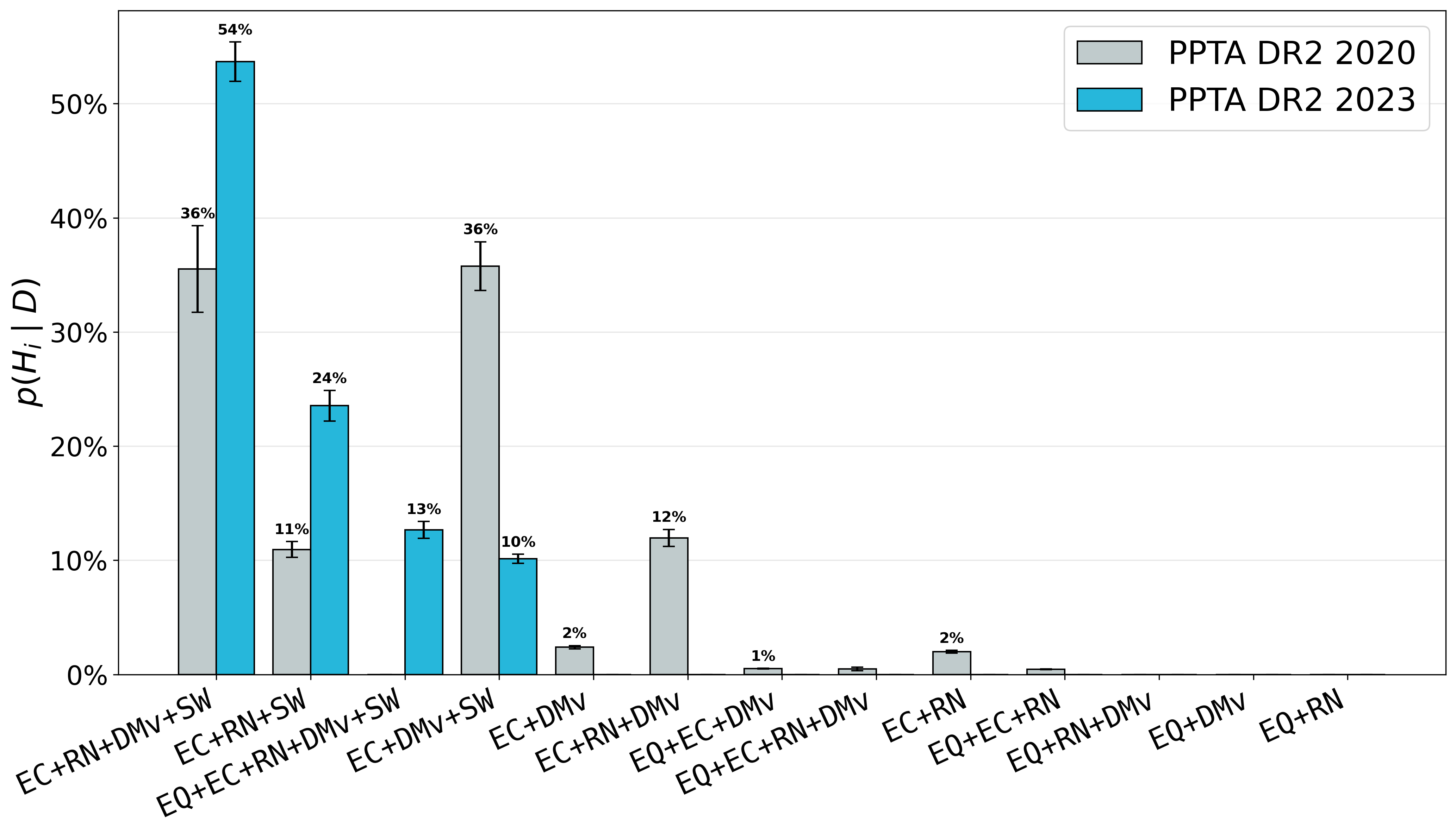}
    {PSR~J2145$-$0750}

  \caption{
    Posterior model probabilities obtained with the 2020~DR2 pipeline
    (grey) and the 2023~DR2 pipeline (teal) for the remaining seven
    pulsars not shown in the main text.
  }
  \label{fig:app_model_2}
\end{figure*}

Figures~\ref{fig:app_model_1} and~\ref{fig:app_model_2} show the posterior
model-probability comparisons under the 2020~DR2 and 2023~DR2 pipelines for
the 19 pulsars not displayed in the main text.

The notation follows that of the main text. EF is present in every model and
is therefore omitted from the abbreviated model labels. EQ, EC, RN, DMv, and
SW denote EQUAD, ECORR, achromatic red noise, dispersion-measure variations,
and solar-wind variations, respectively.

\section{Solar-wind posterior distributions}

\label{app:solar_wind_posteriors}

\begin{figure*}
  \centering

  \pulsarpanel
    {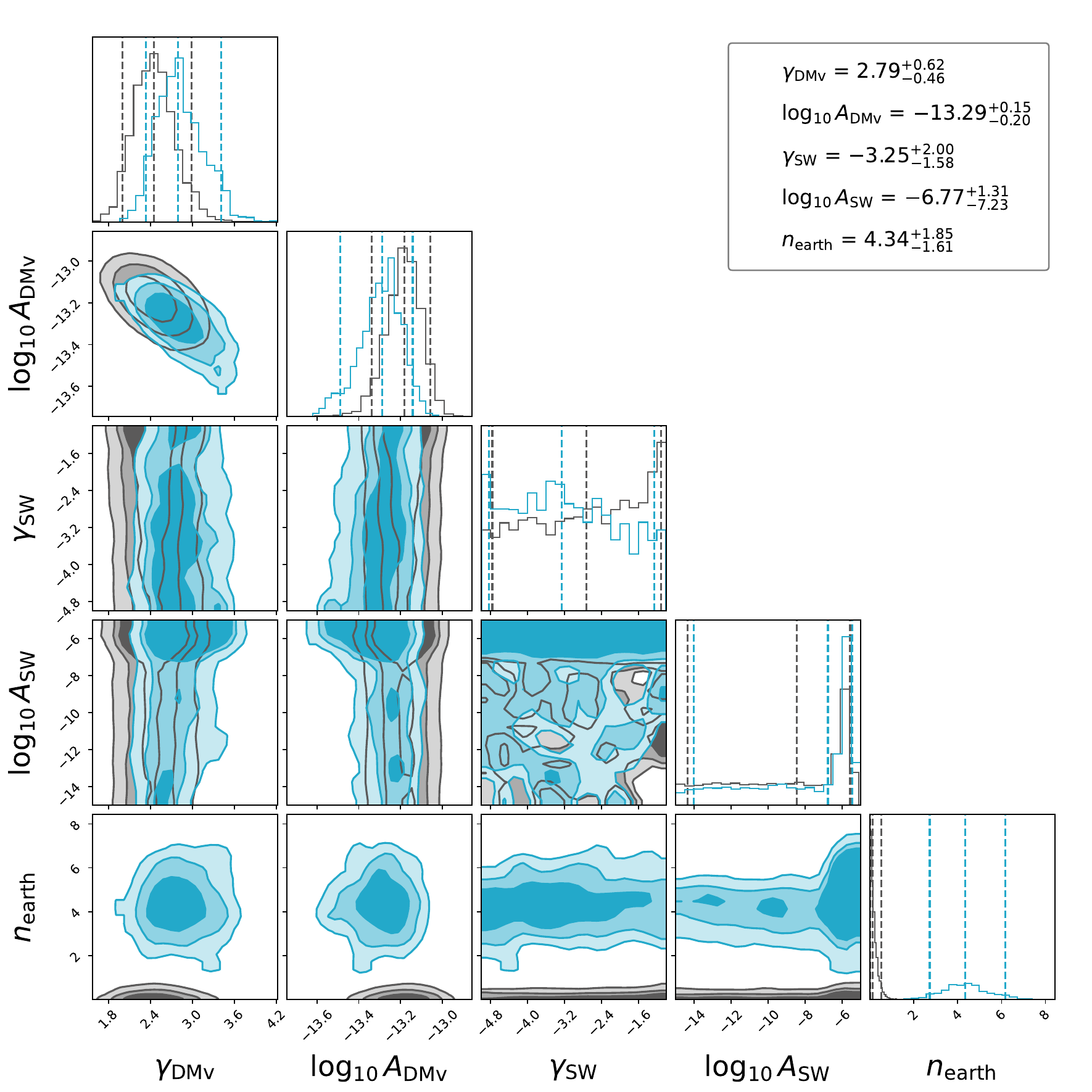}
    {PSR~J1600$-$3053}
  \hfill
  \pulsarpanel
    {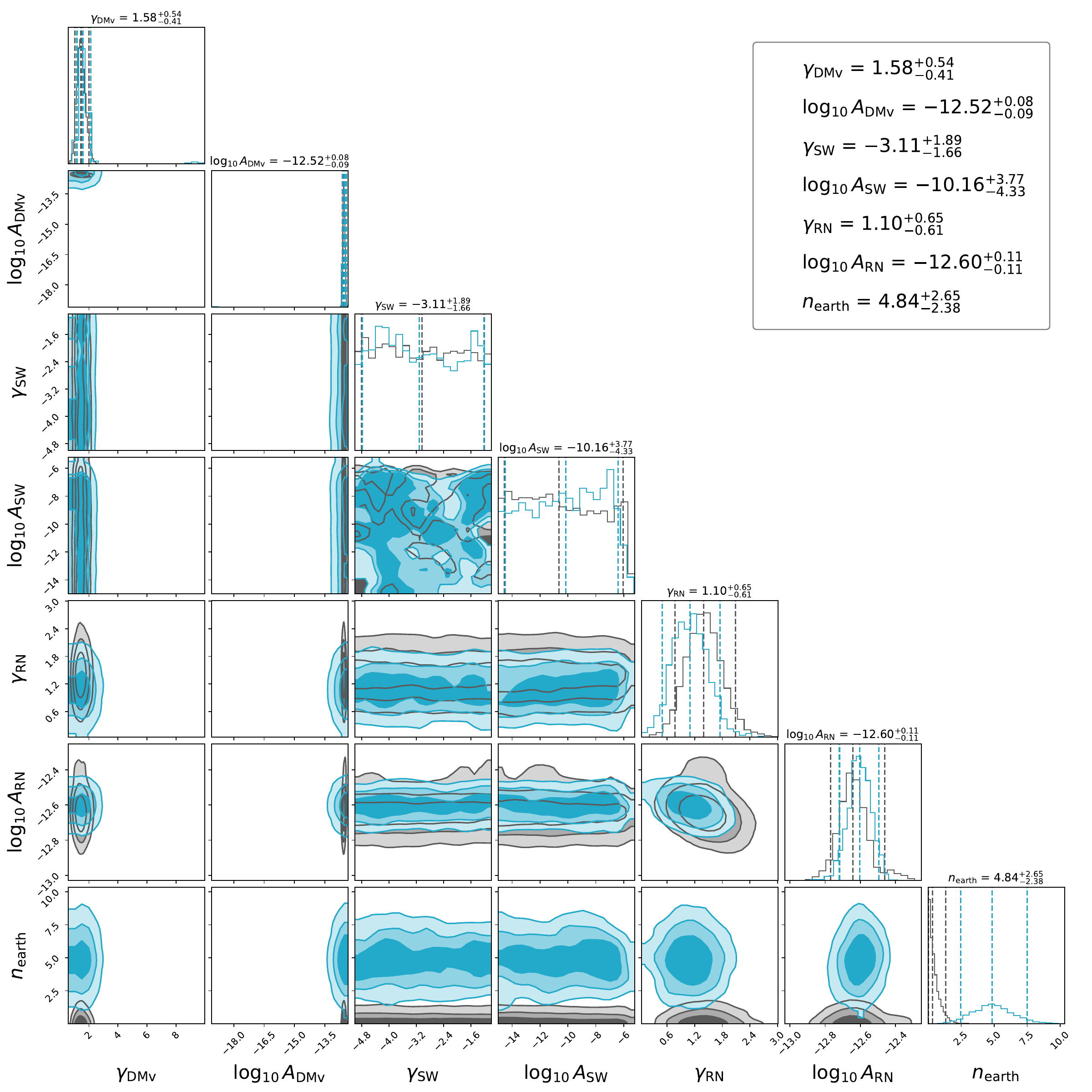}
    {PSR~J1643$-$1224}
  \hfill
  \pulsarpanel
    {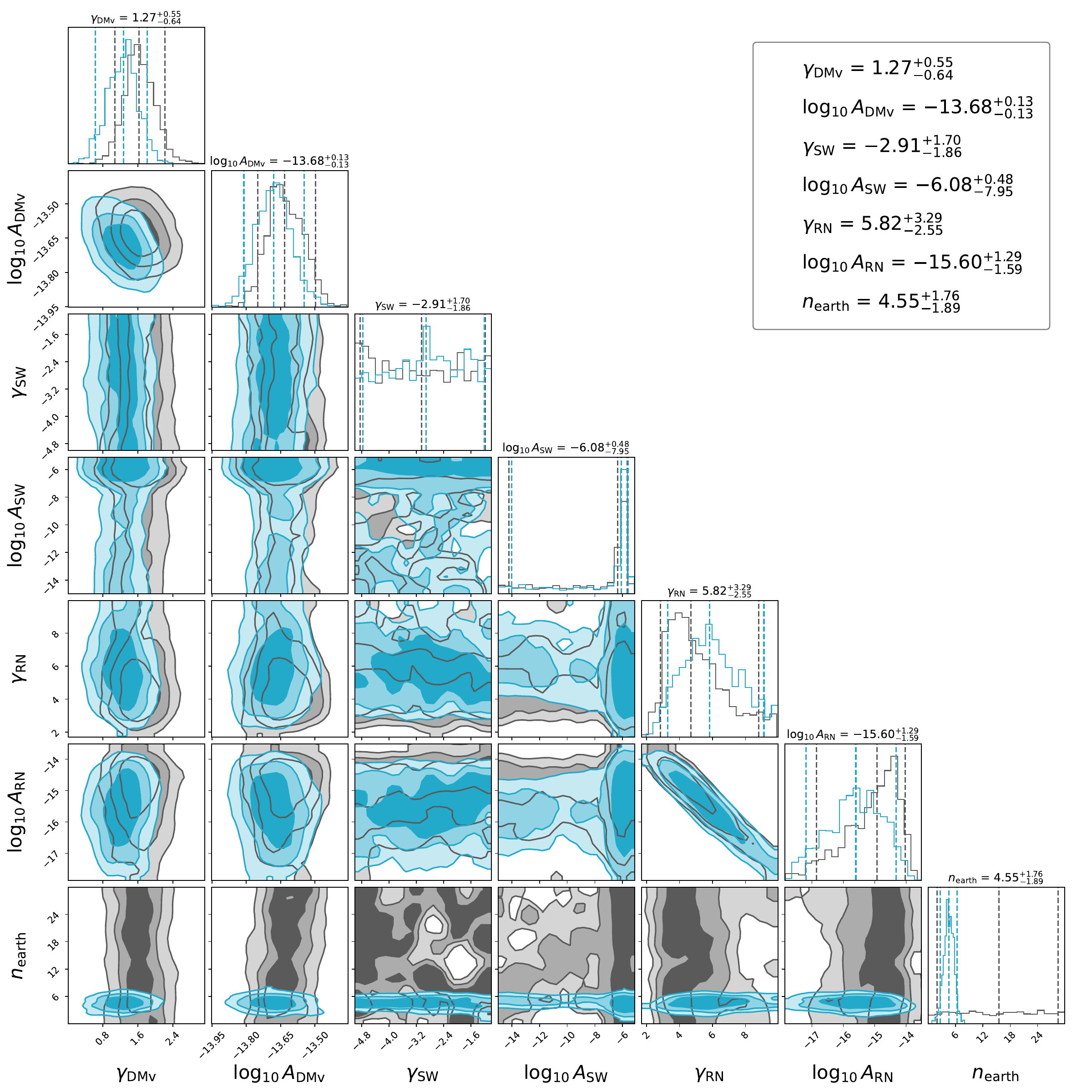}
    {PSR~J1713$+$0747}

  \par\vspace{4pt}

  \pulsarpanel
    {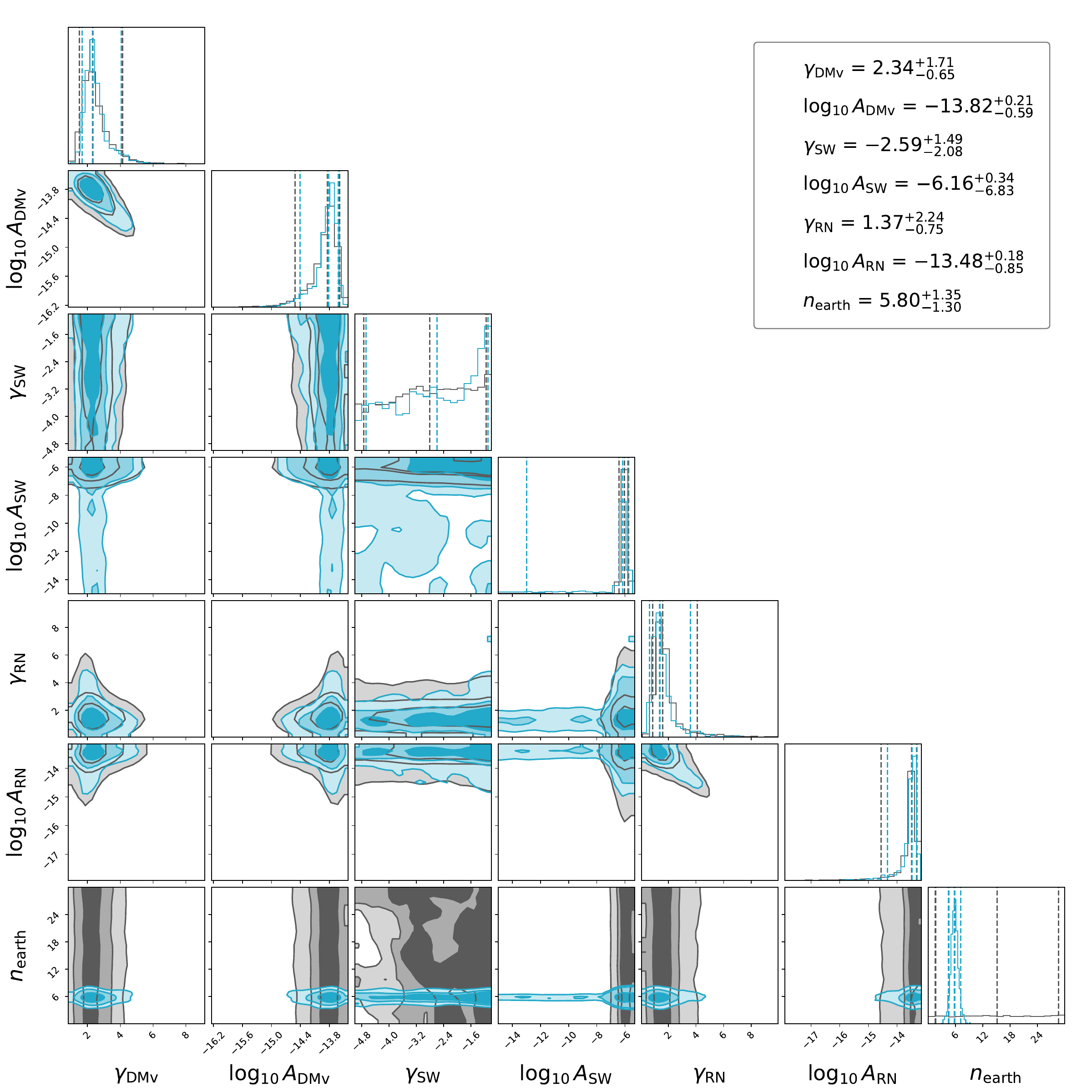}
    {PSR~J1744$-$1134}
  \hfill
  \pulsarpanel
    {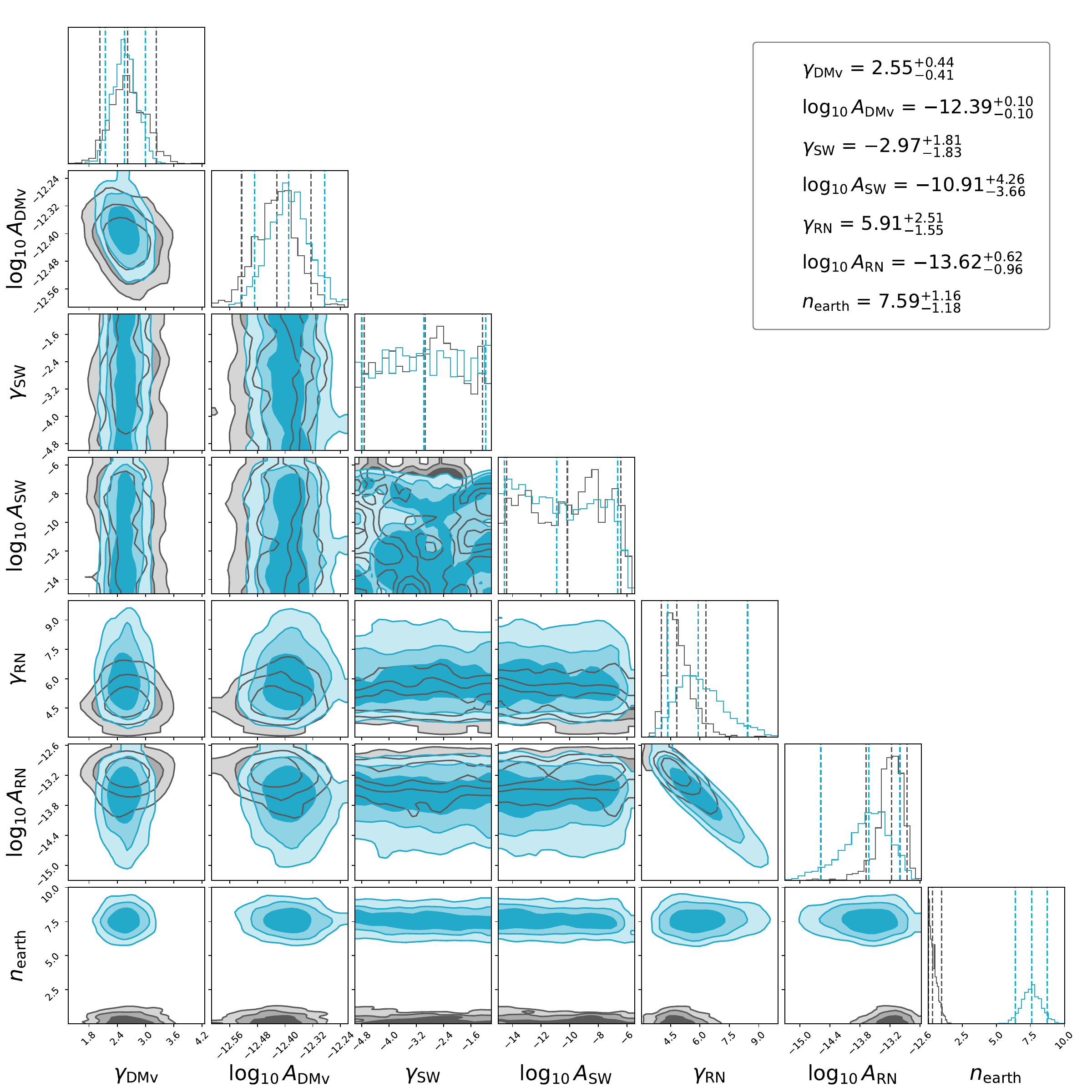}
    {PSR~J1824$-$2452A}
  \hfill
  \pulsarpanel
    {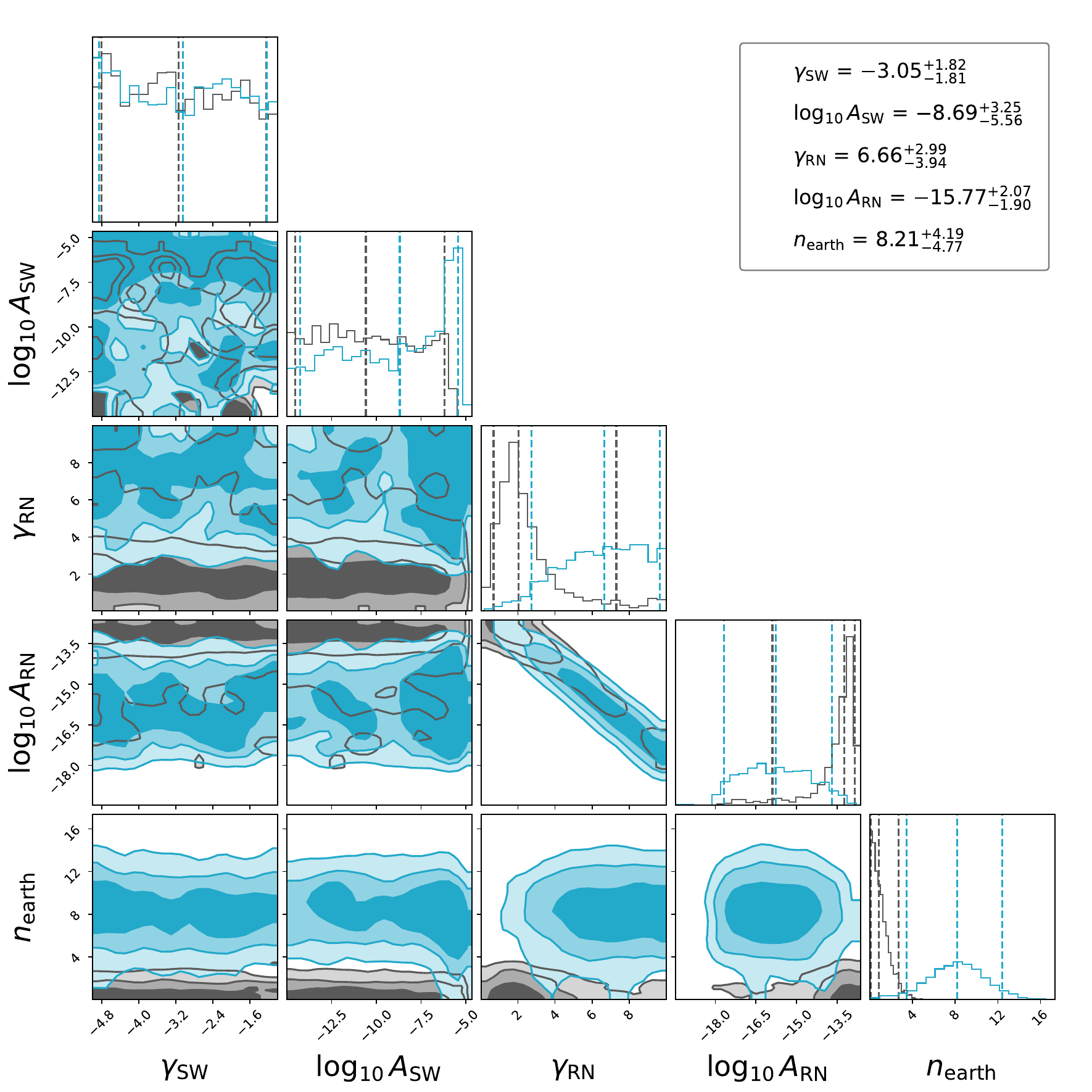}
    {PSR~J2124$-$3358}

  \par\vspace{4pt}

  \pulsarpanel
    {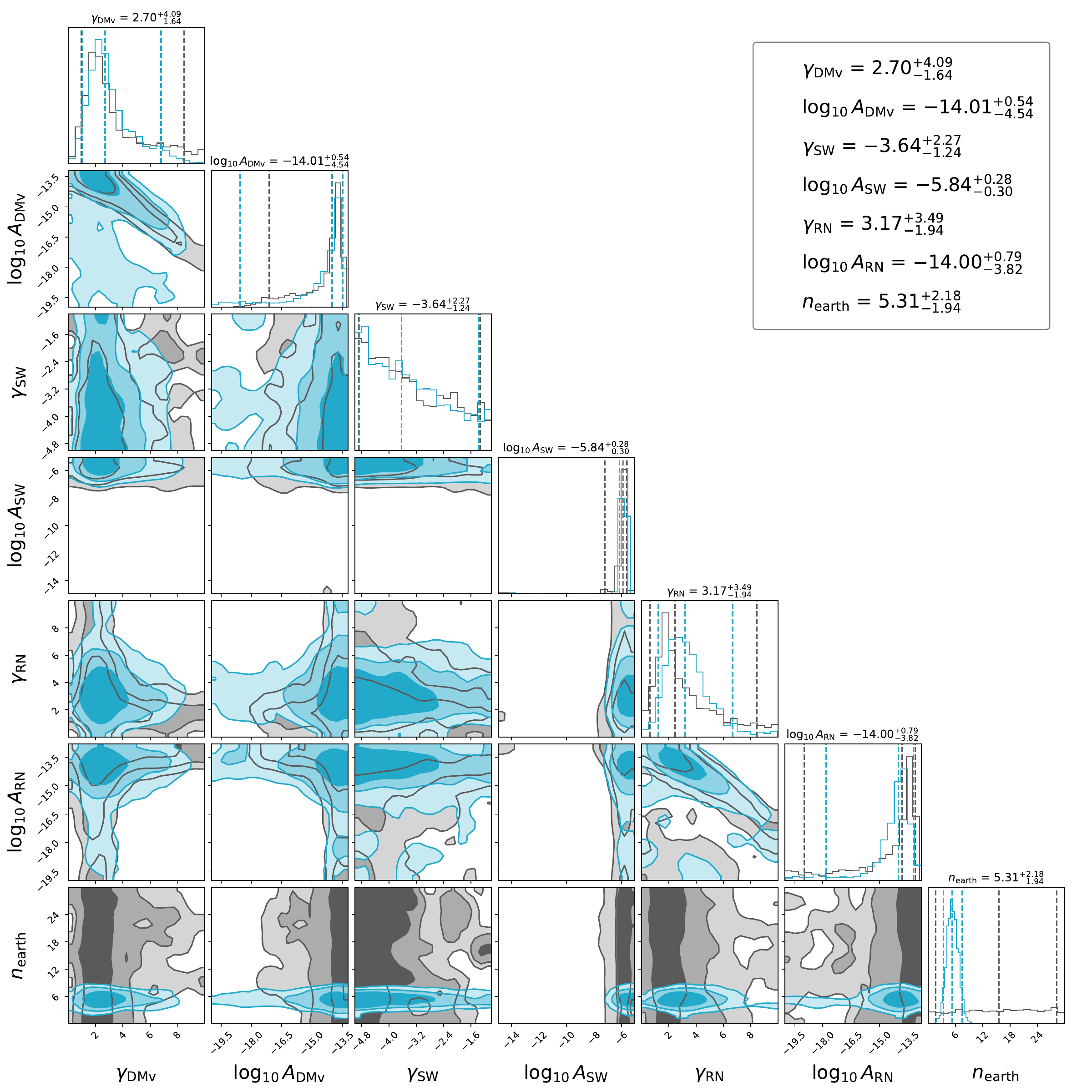}
    {PSR~J2145$-$0750}

  \caption{
    Joint and marginalized posterior distributions for the parameters of
    the preferred noise models of the seven additional pulsars showing
    statistically significant solar-wind variations.  Results obtained with
    the 2020 PPTA DR2 pipeline are shown in grey, while those obtained with
    the 2023 PPTA DR2 pipeline are shown in teal.
  }
  \label{fig:app_corner_sw}
\end{figure*}

Figure~\ref{fig:app_corner_sw} shows the joint and marginalized posterior distributions for the preferred noise models of seven additional pulsars whose highest-posterior model under the 2023 pipeline contains a solar-wind component. 

\clearpage
\bibliographystyle{apsrev4-2}

\bibliography{prd/biblio_prd}

\end{document}